\documentclass{article}
\usepackage[utf8]{inputenc}
\usepackage{amssymb}
\usepackage{amsmath}
\usepackage{amsthm}
\usepackage[margin=0.9in]{geometry}
\usepackage{stackrel}
\usepackage{enumitem}
\usepackage{graphicx}
\usepackage{dsfont}
\usepackage{bm} 
\usepackage{pdfpages}
\usepackage{indentfirst}
\usepackage{setspace}
\usepackage{booktabs}
\usepackage{subcaption}
\usepackage{url}
\usepackage{float}

\usepackage{xcolor}

\definecolor{dark-maroon}{HTML}{5D0F0D}
\definecolor{navyblue}{HTML}{0A3044}

\definecolor{purple}{HTML}{5601A4}
\definecolor{navy}{HTML}{0D3D56}
\definecolor{ruby}{HTML}{9a2515}
\definecolor{alice}{HTML}{107895}
\definecolor{daisy}{HTML}{EBC944}
\definecolor{coral}{HTML}{F26D21}
\definecolor{kelly}{HTML}{829356}
\definecolor{cranberry}{HTML}{E64173}
\definecolor{jet}{HTML}{131516}
\definecolor{asher}{HTML}{555F61}
\definecolor{slate}{HTML}{314F4F}
\definecolor{blue}{HTML}{0000EE}

\usepackage{hyperref}
\hypersetup{
    colorlinks= true,
    citecolor= blue,
    linkcolor= blue,
    filecolor= blue,      
    urlcolor= blue,
}

\usepackage{natbib}
\setcitestyle{comma,aysep={}}

\newcommand{\norm}[1]{\left\lVert#1\right\rVert} 
\newcommand{\convd}{\stackrel{d}{\rightarrow}} 
\newcommand{\plim}{\stackrel{p}{\rightarrow}} 
\global\long\def\expec#1{\mathbb{E}\left[#1\right]}%
\newcommand{\condexpec}[2]{\mathbb{E}\left[#1 \ \vert \ #2\right]}
\newcommand{\T}{\mathcal{T}}
\newcommand{\SSR}{\text{SSR}}

\newcommand{\isum}{\sum_{i = 1}^N}

\newtheorem{lemma}{Lemma}[section]
\newtheorem{theorem}{Theorem}[section]

\newtheorem{cor}{Corollary}[section]

\newtheorem{assumption}{Assumption}[section]
  
\newtheorem{remark}{Remark}[section]
\numberwithin{table}{section}
\numberwithin{figure}{section}

\title{An Averaging Alternative to Pre-Trend Testing
}
\author{Nicholas L. Brown\\{Department of Economics}\\{Florida State University} \and Qiushi Bu\\{Department of Biostatistics}\\{Columbia University} }
\date{\today}

\begin{document}

\maketitle

\begin{abstract}
    We study difference-in-differences (DID) estimation of average treatment effects on the treated when the researcher is uncertain about which pre-treatment periods satisfy the parallel trends assumption. \citet{Roth_2022} shows that pre-trend testing can induce bias, complementing the statistical literature on post-selection inference. We propose the model averaged difference-in-differences (MADID) estimator, a weighted average of the candidate $2\times2$ DID estimators. The weights are normalized exponential functions of the candidate residual sums of squares, so implementation requires no specification test. We derive MADID's asymptotic properties under conditions that make invalid comparisons detectably more variable than at least one valid comparison. Under these conditions, MADID is consistent when the parallel trends assumption holds for at least one pre-treatment period. Although its joint limiting distribution across post-treatment periods is generally non-Gaussian, inference can be conducted by subsampling or simulation.
\end{abstract}

\doublespacing

\newpage

\section{Introduction}

Difference-in-differences (DID) is widely used in the empirical social sciences. Its usual justification in policy evaluation is the parallel trends assumption: absent the intervention, the mean untreated potential outcomes of treated and control units would have evolved in parallel. Recent work studying extensions of DID under parallel trends includes \citet{De_Chaisemartin_d'haultfoeuille_2020}, \citet{goodman-bacon_2021}, \citet{Sun_Abraham_2021}, \citet{Callaway_SantAnna_2021}, \citet{Borusyak_et_al_2024}, and \citet{Wooldridge_2025}, among others. Assessing the credibility of parallel trends is therefore central to empirical practice.

For applied microeconomists, pre-trend testing is a commonly used diagnostic tool for assessing the credibility of the parallel trends assumption. \citet{Roth_2022} documents that empirical studies often interpret statistically insignificant pre-treatment coefficients as evidence in favor of parallel trends. However, Roth shows that such tests can be severely underpowered against empirically relevant deviations from parallel trends and that conditioning subsequent estimation on the outcome of these tests can substantially increase bias. These concerns are closely related to the broader post-selection inference problem studied by \citet{Leeb_Potscher_2005}: conventional inference does not account for the fact that the same data are used both to assess specification validity and to estimate the selected model. Pre-trend testing also does not provide a satisfactory rule for how to proceed when some pre-treatment periods appear more credible than others. Individual tests raise a multiple-testing problem; sequential tests require choices about their order and stopping rule; and a joint rejection does not identify which comparison to use instead. A fixed significance cutoff can also produce sharply different specification choices for similar evidence, such as $p$-values of 0.04 and 0.06. Thus, pre-trend testing leaves unresolved both the choice of comparison periods and the uncertainty induced by that choice.

We address this specification uncertainty by replacing discrete pre-testing decisions with model averaging over the available pre-treatment comparisons. In our setting, a model consists of a two-by-two ($2 \times 2$) DID design between outcomes in a target post-treatment period and a candidate pre-treatment period. We construct the weights using a calibrated information criterion designed to reflect the sampling variation in the candidate residual sums of squares (SSRs). Our model averaged difference-in-differences (MADID) estimator for a given treatment effect is a weighted average of all candidate $2 \times 2$ DID estimators. We show that MADID is consistent for the average treatment effect on the treated (ATT) when all population SSR minimizers identify the ATT. Our factor-matching and variance-separation conditions provide sufficient conditions for this result. The researcher therefore need not classify candidate pre-treatment periods as valid or invalid before estimation.

Model averaging methods can be broadly divided into Bayesian\footnote{See \citet{Steel_2020} for an overview of Bayesian model averaging techniques.} and frequentist approaches. We focus on the frequentist literature, which avoids specifying prior probabilities over candidate models. \citet{Hjort_Claeskens_2003} develop asymptotic theory for frequentist model averaging under local misspecification. \citet{Hansen_2007} selects averaging weights by minimizing a Mallows criterion that balances in-sample fit against model complexity, and establishes asymptotic optimality under squared-error loss for a discrete class of weights under homoskedasticity. \citet{Wan_Zhang_Zou_2010} extend least-squares model averaging to broader collections of linear candidate models, including non-nested specifications. \citet{Hansen_Racine_2012} propose a jackknife model averaging estimator whose asymptotic optimality result allows heteroskedastic errors.

Recent work has examined inferential properties of these frequentist model averaging estimators. \citet{Zhang_Liu_2019_inference} study the Mallows and jackknife model averaging estimators for nested linear models and prove consistency when the true model is included among the candidates. They derive a non-standard limiting distribution in which the weights on underfitted models (i.e., models that exclude relevant covariates) converge to zero. \citet{Wang_et_al_2019} consider information-criterion weights for candidate models estimated by maximum likelihood and obtain a similar non-standard limiting distribution. These results do not apply directly to our DID setting. The candidate models in these analyses typically differ in covariate specification but use a common outcome sample. By contrast, our candidate DID estimators regress different long differences of the outcome on a constant and a treatment indicator, so each candidate uses a different pair of time periods from the same panel. With $T$ fixed, the normalized SSRs fluctuate at the $N^{-1/2}$ rate under random sampling. We therefore scale the criterion by $\sqrt N$ rather than $N$, so that sampling fluctuations remain first-order when population SSRs are equal. Our contribution to the model averaging literature is therefore to develop asymptotic theory for estimators of a common target constructed from potentially misspecified, non-nested time comparisons.

Other model averaging approaches to treatment effect estimation include \citet{Lu_2015}, \citet{Kitagawa_Muris_2016}, and \citet{Firpo_et_al_2025}. These methods address uncertainty over covariate or propensity-score specifications, but assume that a class of correctly specified models are known. MADID instead averages over pre-treatment comparisons, assigns non-negative weights that sum to one by construction, allows some candidate DID estimators to be globally misspecified when all population SSR minimizers identify the ATT, and does not require the researcher to identify the misspecified candidates in advance.

We derive our asymptotic results under an interactive factor model, which provides a flexible representation of non-parallel trends through interactions between time-varying common factors and unit-specific factor loadings. Existing factor-based approaches to treatment effect estimation include quasi-differencing methods \citep{Callaway_Karami_2023,Callaway_Tsyawo_2023,Brown_Butts_2025}, cross-sectional averaging \citep{Brown_et_al_2023}, and least squares estimation of the factors and loadings \citep{Gobillon_Magnac_2016,Xu_2017,Chan_Kwok_2022}. These methods typically require either a sufficiently large time dimension, correct specification of the factor structure, or additional proxy variables. MADID instead operates in a short-panel setting. Our sufficient conditions require a factor-matched pre-treatment period, whose common factors equal those in the target post-treatment period, and variance separation from factor-mismatched comparisons. The separation condition requires the residual-variance increase due to factor mismatch to dominate differences in idiosyncratic covariance across comparisons. The existence of a valid comparison alone does not ensure that the criterion favors valid comparisons.

Several recent papers also construct weights over time periods. \citet{Arkhangelsky_et_al_2021} and \citet{Athey_et_al_2026} combine time weighting with unit weighting or matrix-completion methods, but their asymptotic frameworks generally require both $N$ and $T$ to grow. Our setting is closest to \citet{Schenk_2025}, who also averages $2\times2$ DID estimators across pre-treatment periods. His time-weighted DID uses variance-minimizing weights and also studies bias reduction under common-factor violations. MADID focuses on consistent estimation when some comparisons are invalid, under the stated separation conditions. Our calibrated criterion therefore serves both a weighting and a screening role, asymptotically downweighting periods with larger population SSRs while retaining sampling uncertainty among tied population minimizers.

In summary, our contribution is threefold. First, we recast uncertainty over the appropriate pre-treatment comparison period in DID as a model averaging problem. Second, we introduce a calibrated information criterion that downweights inferior candidates while retaining uncertainty when population criteria are equal. Third, we derive the joint limiting distribution across post-treatment ATTs, accounting for dependence between the DID estimates and the criterion fluctuations that determine their weights. The generally non-Gaussian limit motivates inference via $K$-sample subsampling and simulation.

The paper proceeds as follows. Section 2 describes the factor model and defines the MADID estimator. Section 3 derives its joint asymptotic distribution. Section 4 develops inference procedures. Section 5 compares finite-sample performance with other selection and averaging procedures in Monte Carlo experiments. Section 6 presents applications studying the local labor market impact of Opportunity Zones and the effect of post-\textit{Dobbs} abortion bans on fertility, which differ in sample size and pre-trend patterns. Section 7 concludes. 

\section{Model and Estimator}

We observe a balanced panel of $N$ individuals over $T$ time periods. Some units are treated immediately after period $T_0$ and remain treated until the end of the panel; we generalize our model to allow staggered interventions in Section \ref{section: staggered intervention}. Let $D_i$ be the overall treatment indicator. In general, $\mathds{1}(A)$ is the indicator function that takes value $1$ if $A$ occurs and $0$ otherwise. $N_1 \equiv \isum D_i$ and $N_0 \equiv N - N_1$ are the total number of treated and untreated units, respectively. We denote $y_{i,t}(1)$ and $y_{i,t}(0)$ as the treated and untreated potential outcomes, respectively. The target parameter is the ATT for period $s > T_0$: 
\begin{equation*}
    \tau_s \equiv \condexpec{y_{i,s}(1) - y_{i,s}(0)}{D_i = 1}.
\end{equation*}

Researchers will often estimate this parameter under a parallel trends assumption, which states that in the absence of treatment, the treated and control units' outcomes would have evolved in parallel:
\begin{equation*}
    \condexpec{y_{i,s}(0) - y_{i,r}(0)}{D_i = 1} = \condexpec{y_{i,s}(0) - y_{i,r}(0)}{D_i = 0} \text{ for $s > T_0$ and $r \in \{1,...,T_0\}$.}
\end{equation*}
The population DID estimand is\footnote{We use $\tau_s$ to denote the actual causal parameter while $\tau_{s,r}$ is the population DID regression coefficient between periods $s$ and $r$.}
\begin{equation*}
    \tau_{s,r} \equiv \condexpec{y_{i,s} - y_{i,r}}{D_i = 1} - \condexpec{y_{i,s} - y_{i,r}}{D_i = 0}.
\end{equation*}
Under a general set of assumptions, it is well known that the DID estimand identifies the ATT if the parallel trends assumption holds between periods $s$ and $r$. 

The parallel trends assumption implies that untreated potential outcomes can be written as a two-way additive error model whose idiosyncratic component is mean zero conditional on treatment status \citep{Wooldridge_2025}. We generalize this model to allow for interactions between unobserved heterogeneity: 
\begin{equation}\label{eq: untreated po equation}
    y_{i,t}(0) = c_i + \theta_t + \bm f_t' \bm \gamma_i + u_{i,t},
\end{equation}
where $c_i$ is a unit-specific intercept, $\theta_t$ is a secular time effect, $\bm f_t$ is a $K \times 1$ vector of common factors, $\bm \gamma_i$ is a $K \times 1$ vector of factor loadings, and $\condexpec{u_{i,t}}{c_i, \bm \gamma_i, D_i} = 0$. We treat the factors as fixed and assume that endogeneity is driven by correlation between the loadings and treatment status. The additive effects $c_i + \theta_t$ are a special case of the factor structure, but we nest it explicitly. If the common factors were observed and their pre-treatment variation identified the relevant group-level loadings, a regression allowing group-specific factor effects could recover the untreated group means in post-treatment periods. Otherwise, standard DID will not generally recover the ATT:
\begin{equation*}
    \tau_{s,t} = \tau_s + (\bm f_s - \bm f_t)' (\condexpec{\bm \gamma_i}{D_i = 1} - \condexpec{\bm \gamma_i}{D_i = 0}).
\end{equation*}
The parallel trends assumption holds if $\bm f_s = \bm f_t$ (similar macro trends) or $\bm \gamma_i$ is mean independent of treatment (similar exposure between groups). 

If $\bm \gamma_i$ is mean independent of treatment, any affine combination of the pre-treatment DID estimators would be consistent and one should use \citet{Schenk_2025} to obtain an efficient estimator. We therefore focus on screening pre-treatment periods based on factor similarity. We define the set of time periods whose macro shocks are equal to the given post-treatment period. For each $s > T_0$, let
\begin{gather*}
    \T_s^1 \equiv \{r \in \{1,...,T_0\} \ | \ \bm f_r = \bm f_s \},\\
    \T_s^2 \equiv \{t \in \{1,...,T_0\} \ | \ \bm f_t \neq \bm f_s \}.
\end{gather*}
If we knew which periods were factor-matched ($\bm f_s=\bm f_r$), we could use them to estimate the ATT.\footnote{We generally use `$r$' for factor-matched periods and `$t$' for other candidate periods.} In practice, this set is unknown, and the pre-testing concerns discussed in the introduction make the specification choice difficult.

Rather than test over the pre-treatment periods, we obtain each possible $2\times2$ DID estimator, then assign weights based on the residual variation in each comparison. The DID estimator of the $s$'th ATT using only pre-treatment period $t$ is
\begin{equation}\label{eq: DID estimator}
    \widehat{\tau}_{s,t} \equiv \left( \frac{1}{N_1}\isum D_i(y_{i,s} - y_{i,t}) \right) - \left( \frac{1}{N_0} \isum (1 - D_i) (y_{i,s} - y_{i,t}) \right).
\end{equation}
 For each target ATT, we seek a set of weights $\bm w_s \equiv (w_{s,1},...,w_{s,T_0})'$ where each weight is between 0 and 1 and all weights sum to 1. Our estimator is then
\begin{equation*}
    \widehat{\tau}_s(\bm w_s) \equiv \sum_{t = 1}^{T_0} w_{s,t} \widehat{\tau}_{s,t}.
\end{equation*}

We are particularly interested in finding weights that directly compare the difference of unobserved factors between periods. Traditional information criterion weights are defined with respect to the log-likelihood. It is well known that DID can be obtained by regressing the long-differenced outcomes on an intercept and treatment dummy. We consider the following SSR:
\begin{equation*}
    \widehat{\SSR}_{s,t} \equiv \isum (y_{i,s} - y_{i,t} - \widehat{\gamma}_{s,t} - D_i \widehat{\tau}_{s,t})^2,
\end{equation*}
where $\widehat{\gamma}_{s,t} \equiv \frac{1}{N_0} \isum (1 - D_i) (y_{i,s} - y_{i,t})$ and $\widehat{\tau}_{s,t}$ is the $2 \times 2$ DID estimator between periods $s$ and $t$.\footnote{For a two-period regression with unit and time fixed effects, the level-regression SSR is one half of the long-difference SSR. This common proportionality factor cancels in the weights below.} The Gaussian working negative log-likelihood, after profiling out the variance using $\widehat{\SSR}_{s,t}/N$, is
\begin{equation*}
    \mathcal L(\widehat\gamma_{s,t},\widehat\tau_{s,t})
    =
    C_N+\frac{N}{2}\log(\widehat{\SSR}_{s,t}/N),
\end{equation*}
where $C_N$ is common across candidates. This calculation motivates the criterion; the asymptotic analysis does not assume Gaussian errors.

We propose a modified information criterion for estimator $t$ of target parameter $\tau_s$:
\begin{equation*}
    \text{IC}_{s,t} \equiv \sqrt{N} \log(\widehat{\SSR}_{s,t} / N).
\end{equation*}
We then study weights of the same form as proposed by \citet{Buckland_et_al_1997}:
\begin{equation}\label{eq: MADID weight}
    \widehat{w}_{s,t} \equiv \frac{\exp(-\text{IC}_{s,t}/2)}{\sum_{q = 1}^{T_0}\exp(-\text{IC}_{s,q}/2)}.
\end{equation}
The model averaged difference-in-differences (MADID) estimator is $\widehat{\tau}_{s}(\widehat{\bm w}_s)$, where $\widehat{\bm w}_s \equiv (\widehat{w}_{s,1},...,\widehat{w}_{s,T_0})'$ and each weight is defined by equation \eqref{eq: MADID weight}. First note that our weights are always strictly greater than zero and sum to one. That they sum to one and are asymptotically bounded imply that MADID is consistent when $\bm \gamma_i$ is mean independent of treatment. Convexity implies that MADID will not extrapolate beyond the estimates of each $2 \times 2$ DID estimators. Any discrete selection procedure will obtain a pooled estimator whose value is bounded between 
\begin{equation*}
    \left[ \min_{t \leq T_0}\{ \widehat{\tau}_{s,t} \}\text{, } \max_{t \leq T_0} \{ \widehat{\tau}_{s, t} \} \right].
\end{equation*}
MADID can take any value within this interval but will not extrapolate beyond it, thus making it a natural alternative to selection.

Our weight differs from the standard information criterion weights in two important ways. First, \citet{Wang_et_al_2019} differentiate between the Akaike and Schwarz information criteria in their approach, which assigns a different penalty to model complexity. In our setting, each candidate model contains the same number of parameters, and therefore the two criteria yield identical weights. Second, our criterion scales the goodness-of-fit component by $\sqrt N$ rather than $N$. The likelihood-based analysis of \citet{Wang_et_al_2019} compares candidate models on a common outcome sample. Here the candidates use different long differences. We instead examine the normalized SSRs, whose deviations from their population means are $O_p(N^{-1/2})$ under random sampling. In practice, $\sqrt{N}$-scaling distributes weight more evenly in small samples compared to $N$-scaling, which tends to select a single period, better reflecting the uncertainty inherent in the selection process.\footnote{In one of our two applications, we consider the effect of the \textit{Dobbs v. Jackson Women's Health Organization} Supreme Court decision on fertility. We follow \citet{Dench_et_al_2024} and consider a sample of $N = 37$ states: with $\sqrt{N}$-scaling, our weights are mostly concentrated on the last pre-treatment period, but each period has a non-zero weight. With $N$-scaling, the weights on the other periods are numerically negligible. } 

We specifically choose the long-difference SSR because differencing removes the unit effect and compares factor gaps across periods. An alternative is the group-level regression
\begin{equation*}
    \min_{\gamma,\eta,\delta,\tau}
    \isum\sum_{q\in\{t,s\}}
    \left( y_{i,q}-\gamma-D_i\eta-\mathds{1}(q=s)\delta
    -D_i\mathds{1}(q=s)\tau \right)^2.
\end{equation*}
The coefficient on the interaction is again the DID estimator. If the level outcomes have finite second moments, however, the normalized SSR converges to
\begin{equation*}
    \sum_{q\in\{t,s\}}
    \left( p Var(y_{i,q}\mid D_i=1)+(1-p)Var(y_{i,q}\mid D_i=0)\right),
\end{equation*}
rather than the variance of the long difference. This criterion would select pre-treatment periods by the strength of their factors, rather than similarity with the target outcome. It also retains the covariance between $c_i$ and other terms in the model, which is irrelevant to the DID estimator because $c_i$ is eliminated by differencing.

\section{Asymptotic Analysis}

We first establish the population SSR ranking and the resulting weight limits, then combine the DID and SSR influence functions in a single joint CLT. This yields the joint distribution of all post-treatment MADID estimators. We also discuss extensions of the basic model. Let $\bm y_i = (y_{i,1},...,y_{i,T})'$ be the $T \times 1$ stack of outcomes, with similar definition for the errors $\bm u_i$ and the individual treatment effects $\bm \tau_i$. We use `$\plim$' and `$\convd$' to denote convergence in probability and convergence in distribution, respectively. Limits are taken as $N \rightarrow \infty$ holding $T$ and the factor dimension fixed. Model equalities and conditional moment restrictions are understood to hold almost surely.

Let $\tau_{i,t}\equiv y_{i,t}(1)-y_{i,t}(0)$ be the unit- and time-specific treatment effect. The observed outcome is $y_{i,t}=y_{i,t}(0)+D_i\tau_{i,t}$. For each candidate $(s,t)$, define
\begin{gather*}
    \gamma_{s,t}\equiv\condexpec{y_{i,s}-y_{i,t}}{D_i=0},\qquad
    e_{i,s,t}\equiv y_{i,s}-y_{i,t}-\gamma_{s,t}-D_i\tau_{s,t},\\
    \SSR_{s,t}^*\equiv\isum e_{i,s,t}^2,\qquad
    \rho_{s,t}\equiv\expec{e_{i,s,t}^2}.
\end{gather*}
We start by making the following basic assumptions:
\begin{assumption}\label{asm: unconditional setting}
    \text{}
    \begin{enumerate}[label=(\roman*)]
        \item $(D_i, \bm \gamma_i, \bm u_i, \bm \tau_i)$ is an i.i.d. sequence with finite fourth moments. 
        \item $\tau_{i,t} = 0$ for all $t\leq T_0$. 
        \item $\expec{D_i} \equiv p \in (0, 1)$.
        \item $\rho_{s,t}>0$ for every $s>T_0$ and $t\leq T_0$.
    \end{enumerate}
    \hfill $\blacksquare$
\end{assumption}

Assumption \ref{asm: unconditional setting}(i) gives the cross-sectional WLLN and CLT; fourth moments are needed for the squared-residual fluctuations. We allow limited treatment effect heterogeneity across units and arbitrary heterogeneity over time. Condition (ii) rules out anticipation, (iii) ensures non-negligible treated and control groups, and (iv) makes the logarithmic criteria well-defined with probability approaching one. On the exceptional event that a group is empty or a candidate SSR is zero, the estimator may be assigned an arbitrary value without affecting the asymptotic results.  

We start by deriving the limiting value of the SSRs. The following result is not novel, but we reproduce it for completeness. 
\begin{lemma}\label{lemma:SSR}
    Let $s > T_0$ and $t \leq T_0$. Under Assumption \ref{asm: unconditional setting}, 
    \begin{equation*}
        \frac{1}{N}\widehat{\SSR}_{s,t} = p Var(y_{i,s} - y_{i,t} \ | \ D_i = 1) + (1 - p) Var(y_{i,s} - y_{i,t} \ | \ D_i = 0) + O_p(N^{-1/2}).
    \end{equation*}
    \hfill $\blacksquare$
\end{lemma}

\noindent All proofs are contained in the appendix. In general, IC weights prefer models with smaller SSRs. The point of our analysis is to provide structural assumptions under which the limiting value of the weights provides an estimator with a meaningful interpretation. Note that the SSRs reflect the variation in the treatment effect, the factor structure, and the idiosyncratic error: 
\begin{align*}
    Var(y_{i,s} - y_{i,t} \ | \ D_i)
    &=
    Var(D_i \tau_{i,s} + (\bm f_s - \bm f_t)' \bm \gamma_i + u_{i,s} - u_{i,t} \ | \ D_i).
\end{align*}
The following assumptions will allow us to separate the contribution of each part of the variance:
\begin{assumption}\label{asm: unconditional trends}
    For $s > T_0$,
    \begin{enumerate}[label=(\roman*)]
        \item For $t = 1,...,T_0$ and $d \in \{0,1\}$, $Var(u_{i,t} \mid D_i=d) = \sigma^2$.
        \item For $t = 1,...,T$, $u_{i,t}$ is uncorrelated with $\tau_{i,s}$ conditional on $D_i$.
        \item $\tau_{i,s}$ is uncorrelated with $\bm \gamma_i$ conditional on $D_i$. 
        \item For all $t \in \T_s^2$,
        \begin{equation*}
            (\bm f_s - \bm f_t)' \left( p Var(\bm \gamma_i \ | \ D_i  = 1) + (1- p) Var(\bm \gamma_i \ | \ D_i  = 0) \right) (\bm f_s - \bm f_t) - 2Cov(u_{i,s}, u_{i,t} - u_{i,r}) > 0,
        \end{equation*}
        for all $r \in \T_s^1$.  \hfill $\blacksquare$
    \end{enumerate}
\end{assumption}

Assumption \ref{asm: unconditional trends}(i) imposes a common conditional variance on the pre-treatment idiosyncratic errors. It is not strictly necessary but simplifies interpretation of the remaining conditions. It neither restricts the post-treatment idiosyncratic variance nor implies homoskedasticity of the composite outcome error. Assumption \ref{asm: unconditional trends}(ii) eliminates covariance between treatment effects and idiosyncratic errors; the conditional mean restriction in equation \eqref{eq: untreated po equation} separately eliminates covariance between factor loadings and idiosyncratic errors. Assumption \ref{asm: unconditional trends}(iii) restricts treatment effect heterogeneity but does not rule out correlation between $\bm \gamma_i$ and $D_i$, or `selection on unobservables'. It is automatically satisfied if the effect of treatment is constant across units. \citet{Schenk_2025} makes a similar assumption. This restriction allows us to simplify the variance decomposition. If this covariance is nonzero, its contribution must instead be included in the variance-separation comparison.\footnote{Suppose $\bm \gamma_i$ captures a worker's latent skills. If high-skill workers benefit more from a job training program, we would expect $Cov(\tau_{i,s}, \bm f_t' \bm \gamma_i \mid D_i) > 0$ for positive technology shocks.}

Assumption \ref{asm: unconditional trends}(iv) is a high level assumption that implies that the limiting SSRs for factor-matched comparisons ($\T_s^1$) are systematically lower than for factor-mismatched comparisons ($\T_s^2$). To see how, note that Assumptions \ref{asm: unconditional trends}(i)--(iii) imply that for $s > T_0$ and $t \leq T_0$,
\begin{align*}
    Var(y_{i,s} - y_{i,t} \mid D_i)
    ={}& D_i Var(\tau_{i,s} \mid D_i=1)
    + (\bm f_s - \bm f_t)' Var(\bm \gamma_i \mid D_i) (\bm f_s - \bm f_t)\\
    &+ Var(u_{i,s} \mid D_i) + \sigma^2
    - 2 Cov(u_{i,t},u_{i,s} \mid D_i).
\end{align*}
Let $r \in \T_s^1$ and $t \in \T_s^2$. Then all terms that do not depend on the pre-treatment periods are differenced away:
\begin{align}
    \frac{1}{N} \widehat{\SSR}_{s,t} - \frac{1}{N} \widehat{\SSR}_{s,r} 
    \plim 
    (& \bm f_s - \bm f_t)' \left( p Var(\bm \gamma_i \ | \ D_i  = 1) + (1- p) Var(\bm \gamma_i \ | \ D_i  = 0) \right) (\bm f_s - \bm f_t) \nonumber\\
    -& 2Cov(u_{i,s}, u_{i,t} - u_{i,r}) .\label{eq:SSRT-S}
\end{align}
where the covariance simplification follows from iterated expectations and the fact that the errors have conditional mean zero given $D_i$. Assumption \ref{asm: unconditional trends}(iv) implies that this quantity is positive for any $r \in \T_s^1$ and $t \in \T_s^2$.

We can derive more basic sufficient conditions that imply Assumption \ref{asm: unconditional trends}(iv). It is natural to assume that $p Var(\bm \gamma_i \ | \ D_i = 1) + (1-p) Var(\bm \gamma_i \ | \ D_i = 0)$ is positive definite.\footnote{See Assumption 6 of \citet{Schenk_2025} for a similar condition.} If the idiosyncratic errors are serially uncorrelated, \ref{asm: unconditional trends}(iv) is satisfied automatically. Instead, suppose we allow correlation but impose homoskedasticity on the post-treatment errors as well. The Cauchy-Schwarz inequality gives us
\begin{align*}
    2Cov(u_{i,s}, u_{i,t} - u_{i,r})
    &\leq
    2 \sqrt{\sigma^2 Var(u_{i,t} - u_{i,r})}\\
    &=
    2\sqrt{\sigma^2 (2 \sigma^2 - 2 Cov(u_{i,t}, u_{i,r}))}\\
    &=
    2\sigma^2\sqrt{2(1 - Corr(u_{i,t}, u_{i,r}))}.
\end{align*}
We can then consider assumptions that imply, for both $d=0,1$ and every $t\in\T_s^2$, $r\in\T_s^1$, 
\begin{equation*}
    (\bm f_s - \bm f_t)' Var(\bm \gamma_i \ | \ D_i = d) (\bm f_s - \bm f_t) > 2 \sigma^2 \sqrt{2(1 - Corr(u_{i,t}, u_{i,r}))}.
\end{equation*}
This sufficient bound is conservative; it is more informative when the errors are strongly positively correlated and the factor gaps are larger. For a single factor and errors with common variance $\sigma^2$ and AR(1) correlation coefficient $\rho$, a sufficient bound is
\begin{equation*}
    \overline V_\gamma>
    \frac{2\sigma^2\sqrt{2(1-\rho^{|t-r|})}}{(f_s-f_t)^2},
    \qquad
    \overline V_\gamma\equiv p Var(\gamma_i\mid D_i=1)+(1-p)Var(\gamma_i\mid D_i=0),
\end{equation*}
for every $t\in\T_s^2$ and $r\in\T_s^1$. If $(f_s-f_t)^2=1$, $|t-r|=1$, and $\rho=0.9$, the sufficient lower bound is approximately $0.894\sigma^2$. A linear factor path $f_t=t$ has no factor-matched pre-treatment period; with a nonzero treated--control loading mean difference, every candidate DID is biased.

We can now present the main results used to derive the asymptotic distribution of $\widehat{\tau}_s(\widehat{\bm w}_s)$. First, we show that our method asymptotically puts zero weight on periods in $\T_s^2$. 
\begin{lemma}\label{lemma: unconditional bad model weights}
    Suppose that $\T_{s}^1$ is non-empty for $s > T_0$ and let $t \in \T_s^2$. Under Assumptions \ref{asm: unconditional setting} and \ref{asm: unconditional trends},
    \begin{equation*}
        \widehat{w}_{s,t} = o_p(N^{-1/2}).
    \end{equation*}
    \hfill $\blacksquare$
\end{lemma}
Lemma \ref{lemma: unconditional bad model weights} states that the weights associated with models in $\T_s^2$ converge to zero fast enough not to affect the limiting distribution of the MADID estimator. For models in $\T_s^1$, we distinguish between SSRs that attain the smallest probability limit and those that do not.
\begin{lemma}\label{lemma: unconditional good model weights}
    Suppose that $\T_s^1$ is non-empty for $s > T_0$ and that Assumptions \ref{asm: unconditional setting} and \ref{asm: unconditional trends} hold. Let $\rho_s^*\equiv\min_{r\in\T_s^1}\rho_{s,r}$ and $\T_s^{1,*}\equiv\{r\in\T_s^1\mid\rho_{s,r}=\rho_s^*\}$. For all $t \in \T_s^1 \setminus \T_s^{1,*}$,
    \begin{equation*}
        \widehat{w}_{s,t} = o_p(N^{-1/2}).
    \end{equation*}
    Let $\bm Z_s \equiv (Z_{s,r})_{r \in \T_s^{1,*}} \sim N(\bm 0, \bm \Xi_s)$, where
    \begin{equation*}
        [\bm \Xi_s]_{r,q} \equiv \frac{Cov(N^{-1/2}\SSR_{s,r}^*, N^{-1/2}\SSR_{s,q}^*)}{(\rho_s^*)^2}, \qquad r,q \in \T_s^{1,*}.
    \end{equation*}
    Then, for all $r \in \T_s^{1,*}$,
    \begin{equation*}
        \widehat{w}_{s,r} \convd w_{s,r}^* \equiv \frac{\exp(-Z_{s,r}/2)}{\sum_{t \in \T_s^{1,*}}\exp(-Z_{s,t}/2)}.
    \end{equation*}
    \hfill $\blacksquare$
\end{lemma}

\begin{remark}[Characterizing $\T_s^{1,*}$]
    For all $r \in \T_s^1$, the limiting SSRs are 
    \begin{equation*}
        \frac{1}{N}\SSR_{s,r}^* \plim p Var(\tau_{i,s} \ | \ D_i = 1) + \sigma^2 + Var(u_{i,s}) - 2 Cov(u_{i,s}, u_{i,r}).
    \end{equation*}
    The set $\T_s^{1,*}$ consists of the factor-matched periods with the largest value of $Cov(u_{i,s},u_{i,r})$. If this covariance has a unique maximizer, then $\T_s^{1,*}$ is a singleton and the limiting weight on that period equals one. If several valid periods attain the maximum, their limiting weights can remain random and reflect first-order sampling variation in the corresponding SSRs. This requires $Var(Z_{s,r}-Z_{s,q})>0$ for some $r,q\in\T_s^{1,*}$; if every such contrast has zero variance, the limiting weights are uniform.

    \hfill $\blacksquare$
\end{remark}

We now have all of the results necessary to derive the joint limiting distribution of the post-treatment MADID estimators. Let
\begin{equation*}
    \mathcal S \equiv \{T_0 + 1,\ldots,T\}
\end{equation*}
denote the set of post-treatment periods. For each $s \in \mathcal S$, define the $T_0 \times T$ differencing matrix $\bm W_s$ that subtracts each pre-treatment outcome from period $s$. We can rewrite the $T_0 \times 1$ stack of $2 \times 2$ DID estimators of $\tau_s$ as
\begin{equation*}
    \frac{1}{N}\isum \left( \frac{D_i}{N_1 / N} - \frac{(1-D_i)}{N_0 / N} \right) \bm W_s \bm y_i.
\end{equation*}
We use the fact that the $r$'th DID estimator always converges to its population linear projection coefficient $\tau_{s,r}$. When $r \in \T_s^1$, $\tau_{s,r} = \tau_s$, the ATT. Define the centered DID influence function for period $s$ as
\begin{equation*}
    \bm \psi_{i,s}
    \equiv
    \frac{D_i}{p}
    \left\{ \bm W_s \bm y_i - \condexpec{\bm W_s \bm y_i}{D_i = 1} \right\}
    -
    \frac{1-D_i}{1-p}
    \left\{ \bm W_s \bm y_i - \condexpec{\bm W_s \bm y_i}{D_i = 0} \right\}.
\end{equation*}
For $r \in \T_s^{1,*}$, let
\begin{equation*}
    \xi_{i,s,r}
    \equiv
    \frac{e_{i,s,r}^2 - \rho_s^*}{\rho_s^*},
\end{equation*}
and stack these terms as $\bm \xi_{i,s} \equiv (\xi_{i,s,r})_{r \in \T_s^{1,*}}$, using the increasing order of the elements of $\T_s^{1,*}$. Stack over post-treatment periods according to
\begin{equation*}
    \bm \psi_i \equiv (\bm \psi_{i,s}')_{s \in \mathcal S}',
    \qquad
    \bm \xi_i \equiv (\bm \xi_{i,s}')_{s \in \mathcal S}',
    \qquad
    \bm V_i \equiv
    \begin{pmatrix}
        \bm \psi_i\\
        \bm \xi_i
    \end{pmatrix}.
\end{equation*}
The covariance matrix of the stacked influence function is
\begin{equation*}
    \bm \Sigma
    \equiv
    Var(\bm V_i)
    =
    \expec{\bm V_i \bm V_i'}
    =
    \begin{pmatrix}
        \bm \Omega & \bm C\\
        \bm C' & \bm \Xi
    \end{pmatrix},
\end{equation*}
where the $(s,s')$ blocks are
\begin{equation*}
    \bm \Omega_{s,s'} \equiv \expec{\bm \psi_{i,s} \bm \psi_{i,s'}'},
    \qquad
    \bm \Xi_{s,s'} \equiv \expec{\bm \xi_{i,s} \bm \xi_{i,s'}'},
    \qquad
    \bm C_{s,s'} \equiv \expec{\bm \psi_{i,s} \bm \xi_{i,s'}'}.
\end{equation*}
Thus $\bm \Sigma$ includes both within-period and cross-period covariance matrices. In general, the cross-period blocks are not zero because the estimators use the same individuals and share pre-treatment outcomes. 
\begin{theorem}\label{theorem: unconditional asymptotic distribution}
    Suppose that $\T_s^{1,*}$ is non-empty for every $s \in \mathcal S$ and that Assumptions \ref{asm: unconditional setting} and \ref{asm: unconditional trends} hold. Let
    \begin{equation*}
        \bm \eta \equiv (\bm \eta_s')_{s \in \mathcal S}',
        \qquad
        \bm Z \equiv (\bm Z_s')_{s \in \mathcal S}',
    \end{equation*}
    where $\bm \eta_s \sim N(\bm 0, \bm \Omega_s)$ for each $s \in \mathcal{S}$ and
    \begin{equation*}
        \begin{pmatrix}
            \bm \eta\\
            \bm Z
        \end{pmatrix}
        \sim N(\bm 0, \bm \Sigma),
    \end{equation*}
    where $\bm \Sigma \equiv Var(\bm V_i)$ as defined above. For each $s \in \mathcal S$, define the $T_0 \times 1$ vector $\bm w_s^*$ by
    \begin{equation*}
        w_{s,r}^*
        \equiv
        \begin{cases}
            \displaystyle\frac{\exp(-Z_{s,r}/2)}
            {\sum_{q\in\T_s^{1,*}}\exp(-Z_{s,q}/2)}, & r\in\T_s^{1,*},\\[6pt]
            0, & r\notin\T_s^{1,*},
        \end{cases}
        \qquad r = 1,\ldots,T_0.
    \end{equation*}
    Then
    \begin{equation*}
        \sqrt{N}
        \begin{pmatrix}
            \widehat{\tau}_{T_0+1}(\widehat{\bm w}_{T_0+1}) - \tau_{T_0+1}\\
            \vdots\\
            \widehat{\tau}_T(\widehat{\bm w}_T) - \tau_T
        \end{pmatrix}
        \convd
        \begin{pmatrix}
            \bm w_{T_0+1}^{*\prime}\bm \eta_{T_0+1}\\
            \vdots\\
            \bm w_T^{*\prime}\bm \eta_T
        \end{pmatrix}.
    \end{equation*}

    \hfill $\blacksquare$

\end{theorem}

Theorem \ref{theorem: unconditional asymptotic distribution} implies consistency and $\sqrt N$-tightness for the vector of post-treatment ATTs. The factor-matching and separation conditions are sufficient: the same proof applies more generally whenever every population SSR minimizer identifies its target ATT. The practitioner need not identify these minimizers before estimation. The underlying vector $(\bm \eta',\bm Z')'$ is jointly Gaussian, but the final vector of MADID estimators is generally not multivariate normal because each $\bm w_s^*$ is a nonlinear softmax function of $\bm Z_s$. The form is similar to the nonstandard model-averaging limits in \citet{Wang_et_al_2019}, where random weights remain on overfitted candidate models.

\subsection{Staggered Intervention}\label{section: staggered intervention}

Adapting our approach to a staggered treatment design is straightforward because we target outcome parameters individually. Suppose the treated population is partitioned into a set of $G$ groups that define when they were treated. Let $\mathcal{G} \subset \{2,...,T\}$ be the set of periods when any unit receives treatment and define $G_{g,i} = \mathds{1}(\text{unit $i$ is treated first at time $g$})$. The relevant ATT is now a group-time specific parameter: 
\begin{equation*}
    \tau_{g,s} = \condexpec{y_{i,s}(g) - y_{i,s}(0)}{G_{g,i} = 1},
\end{equation*}
where $y_{i,s}(0)$ is still the untreated potential outcome but $y_{i,s}(g)$ is the treated potential outcome at time $s$ for a unit treated at time $g$. 

\citet{Callaway_SantAnna_2021} note that the appropriate parallel trends assumption is not unique when treatment adoption is staggered. For example, one may assume parallel trends between each treated cohort and the never-treated group:
\begin{equation*}
    \condexpec{ y_{i,s}(0) - y_{i,s-1}(0)}{G_{g,i} = 1} = \condexpec{ y_{i,s}(0) - y_{i,s-1}(0)}{D_i = 0} \text{ for all $s$ and all $g \in \mathcal{G}$.}
\end{equation*}
Alternatively, one may use units that have not yet been treated by the target period. Let $C_{g,i,s}=1$ indicate that unit $i$ is an admissible control for cohort $g$ at period $s$ under the chosen comparison-group definition, with $G_{g,i}C_{g,i,s}=0$. Hold this group fixed across candidate base periods. For a target period $s\geq g$ and a base period $t<g$, the corresponding identifying restriction is
\begin{equation*}
    \condexpec{y_{i,s}(0)-y_{i,t}(0)}{G_{g,i}=1}
    =
    \condexpec{y_{i,s}(0)-y_{i,t}(0)}{C_{g,i,s}=1}.
\end{equation*}
This notation accommodates either never-treated or not-yet-treated controls. Given the chosen comparison group, the relevant DID estimator is
\begin{equation*}
    \widehat{\tau}_{g,s,t} = \left( \frac{1}{\isum G_{g,i}} \isum G_{g,i} (y_{i,s} - y_{i,t}) \right) - \left( \frac{1}{\isum C_{g,i,s}} \isum C_{g,i,s}(y_{i,s} - y_{i,t}) \right).
\end{equation*}
To construct the SSR, we must also define 
\begin{equation*}
    \widehat{\gamma}_{g,s,t} = \frac{1}{\isum C_{g,i,s}} \isum C_{g,i,s}(y_{i,s} - y_{i,t}).
\end{equation*}
We can then create our weights using the following SSRs: 
\begin{equation*}
    \widehat{\SSR}_{g,s,t} = \isum (G_{g,i} + C_{g,i,s}) (y_{i,s} - y_{i,t} - \widehat{\gamma}_{g,s,t} - G_{g,i} \widehat{\tau}_{g,s,t})^2.
\end{equation*}
Let $N_{g,s}\equiv\isum(G_{g,i}+C_{g,i,s})$. We then construct the weights as
\begin{gather*}
    \text{IC}_{g,s,t}\equiv\sqrt{N_{g,s}}\log(\widehat{\SSR}_{g,s,t}/N_{g,s}),\\
    \widehat{w}_{g,s,t}\equiv
    \frac{\exp(-\text{IC}_{g,s,t}/2)}
    {\sum_{q=1}^{g-1}\exp(-\text{IC}_{g,s,q}/2)},
    \qquad t=1,\ldots,g-1.
\end{gather*}

The weights and inference procedures can then be constructed cohort by cohort and target period by target period as in the common-adoption design. The selected sample must have positive cohort and control proportions, finite fourth moments, positive residual variances, and the corresponding conditional mean, factor-matching, and SSR-separation conditions. These conditions must hold given cohort and control-group membership, not merely overall treatment status. Joint inference must retain covariance induced by shared units.

\section{Inference}

This section discusses inference of the ATT using the limiting result in Theorem \ref{theorem: unconditional asymptotic distribution}. We propose three methods: the first uses a studentization approach that requires assumptions beyond those needed for the convergence results. The second and third do not require this condition, but rely on either subsampling or selection via an asymptotic rate strictly slower than $\sqrt{N}$. 

\subsection{Studentization}\label{section: studentization}

Theorem \ref{theorem: unconditional asymptotic distribution} shows that the MADID estimator's weights can remain random when several periods attain the minimum population SSR. For a fixed $s$, the relevant marginal block of the joint Gaussian limit is
\begin{equation*}
    \begin{pmatrix}
        \bm \eta_s\\
        \bm Z_s
    \end{pmatrix}
    \sim
    N\left(
        \bm 0,
        \begin{pmatrix}
            \bm \Omega_s & \bm C_s\\
            \bm C_s' & \bm \Xi_s
        \end{pmatrix}
    \right),
\end{equation*}
where $\bm \Omega_s \equiv \bm \Omega_{s,s} = \expec{\bm \psi_{i,s} \bm \psi_{i,s}'}$, $\bm C_s \equiv \bm C_{s,s}$, and $\bm \Xi_s \equiv \bm \Xi_{s,s}$. Joint inference across post-treatment periods instead uses the full matrix $\bm \Sigma$. We can characterize the conditional distribution of the estimator, which provides insight into when traditional inference will work. 
\begin{cor}\label{cor: eta given Z}
    Suppose the conditions of Theorem \ref{theorem: unconditional asymptotic distribution} hold with at least two periods in $\T_s^{1,*}$, and assume that $\bm\Xi_s$ is positive definite. Then the distribution of $\bm w_s^{*\prime}\bm \eta_s$ conditional on $\bm Z_s$ is
    \begin{equation*}
        N\left( \bm w_s^{*\prime} \bm C_s \bm \Xi_s^{-1} \bm Z_s, \bm w_s^{*\prime} \bm \Omega_s \bm w_s^* - \bm w_s^{*\prime} \bm C_s \bm \Xi_s^{-1} \bm C_s' \bm w_s^* \right),
    \end{equation*}
    where $\bm C_s \equiv Cov(\bm \eta_s, \bm Z_s)$.
    \hfill $\blacksquare$
    
\end{cor}

The corollary uses the fact that the conditional distribution of $\bm \eta_s$ given $\bm Z_s$ is 
\begin{equation*}
    \bm \eta_s | \bm Z_s \sim N(\bm C_s \bm \Xi_s^{-1} \bm Z_s, \bm \Omega_s - \bm C_s \bm \Xi_s^{-1} \bm C_s'),
\end{equation*}
using the standard conditional-normal formula.\footnote{See Section 2.5.1 of \citet{Anderson_1984}.} If $\bm \eta_s$ and $\bm Z_s$ are asymptotically uncorrelated, their joint Gaussian limit makes them independent, and the conditional distribution of $\bm w_s^{*\prime}\bm \eta_s$ given $\bm Z_s$ is $N(\bm 0,\bm w_s^{*\prime}\bm \Omega_s\bm w_s^*)$. The full $T_0\times T_0$ matrix $\bm \Omega_s$ can be estimated without knowing $\T_s^{1,*}$ by
\begin{equation*}
    \widehat{\bm \Omega}_s = \frac{1}{N} \isum \left( \frac{D_i}{N_1 /N} - \frac{(1 - D_i)}{ N_0 / N} \right)^2 \left( \bm W_s \bm y_i - \widehat{\bm \gamma}_s - D_i \widehat{\bm \tau}_s  \right) \left( \bm W_s \bm y_i - \widehat{\bm \gamma}_s - D_i \widehat{\bm \tau}_s  \right)',
\end{equation*}
where $\widehat{\bm \gamma}_s$ and $\widehat{\bm \tau}_s$ are the $T_0 \times 1$ vectors of intercept and DID estimators for the candidate models, respectively. Under the maintained moment conditions, $\widehat{\bm \Omega}_s\plim\bm \Omega_s$. When $\widehat{\bm w}_s$ has a nondegenerate random limit, the quadratic form does not generally converge in probability to a constant; instead, it converges jointly in distribution to $\bm w_s^{*\prime}\bm \Omega_s\bm w_s^*$. The next lemma gives sufficient conditions under which this random normalization nevertheless yields a standard normal limit.

\begin{lemma}\label{lemma: eta and Z independent}
    Suppose the conditions of Theorem \ref{theorem: unconditional asymptotic distribution} hold and the submatrix $\bm\Omega_s^{**}\equiv([\bm\Omega_s]_{r,q})_{r,q\in\T_s^{1,*}}$ is positive definite. Let $\bm v_{i,s}\equiv(u_{i,s}-u_{i,t})_{t\in\T_s^{1,*}}$. Further, suppose that
    \begin{enumerate}[label=(\roman*)]
        \item $\bm v_{i,s}$ is independent of $D_i$.
        \item $\bm v_{i,s}$ is independent of $\tau_{i,s}$ conditional on $D_i = 1$.
        \item $\condexpec{(\tau_{i,s} - \tau_s)^3}{D_i = 1} = 0$.  
    \end{enumerate}
    Then $\bm w_s^{*\prime} \bm C_s = \bm 0$ and
    \begin{equation*}
        \frac{\sqrt{N}(\widehat{\tau}_s(\widehat{\bm w}_s) - \tau_s)}{\sqrt{\widehat{\bm w}_s' \widehat{\bm \Omega}_s \widehat{\bm w}_s}} \convd N(0, 1),
    \end{equation*}
    \hfill $\blacksquare$
    
\end{lemma}
The limiting weights are driven by the SSR fluctuations for periods in $\T_s^{1,*}$, whereas $\bm \eta_s$ is driven by the DID influence functions. Lemma \ref{lemma: eta and Z independent} therefore imposes restrictions that eliminate the relevant third-order cross-moments. Conditions (i) and (ii) concern the joint vector of idiosyncratic differences, not merely each component separately. Condition (iii) rules out skewness in the treated effect distribution and may be restrictive. For example, a policy maker may create a job training program that benefits all eligible workers, but particularly lower-skilled workers. When these conditions hold, however, conventional standard-normal critical values are valid after studentization. 

\subsection{Subsampling}\label{section:subsampling}
We now turn to obtaining valid inference without requiring $\bm C_s = \bm 0$ for any $s \in \mathcal{S}$. We propose using simulated methods and subsampling; both methods require using an object with a slower rate of convergence than $\sqrt{N}$ because of the normalization used to construct $\widehat{\bm w}_s$. To see why the usual nonparametric bootstrap can fail when population criteria are tied, let $\widehat{\SSR}_{s,r}^{(b)}$ be a realization of the SSR given a bootstrap sample $b$. After removing the common factor involving $\rho_s^*$, the numerator of $\widehat{w}_{s,r}^{(b)}$ can be written as
\begin{align*}
    &\exp\left\{ -\frac{\sqrt{N}}{2}\left( \log(\widehat{\SSR}_{s,r}^{(b)} / N) - \log(\rho_s^*) \right) \right\}\\
    &\qquad= \exp\left\{ -\frac{\sqrt{N}}{2}\left[ \log(\widehat{\SSR}_{s,r}^{(b)} / N) - \log(\widehat{\SSR}_{s,r} / N)
    + \log(\widehat{\SSR}_{s,r} / N) - \log(\rho_s^*) \right] \right\}.
\end{align*}
Even if the bootstrap properly recovers the limiting distribution of the centered sample SSR, the additional term has an $O_p(1)$ contribution after $\sqrt N$ scaling. Its contrasts across tied candidates can shift the conditional bootstrap weight distribution.

We instead apply the subsampling method of \citet{Politis_Romano_1994}. Because DID is a difference between treated and control means, we use the $K$-sample construction of \citet{Politis_Romano_2010} and treat the two groups as independent samples. Let $N_d=\sum_{i=1}^N\mathds{1}(D_i=d)$ for $d\in\{0,1\}$. Choose integers $b_{d,N}$ such that $b_{d,N}\rightarrow\infty$ and $b_{d,N}/N_d\rightarrow0$, and write $b_N=b_{0,N}+b_{1,N}$. For subsample draw $j$, independently select $b_{1,N}$ treated units and $b_{0,N}$ control units without replacement, and recompute all candidate DID estimators, SSRs, MADID weights, and the resulting estimator $\widehat\tau_{s,b_N}^{(j)}$ within that strict subset, using $\sqrt{b_N}$ in the criterion.

Given $B$ independent random strict subsets conditional on the observed panel, define the empirical subsampling distribution
\begin{equation*}
    \widehat L_{N,B}(x)
    =\frac{1}{B}\sum_{j=1}^B
    \mathds{1}\left\{
    \sqrt{b_N}
    \left(\widehat\tau_{s,b_N}^{(j)}-\widehat\tau_s\right)
    \leq x\right\}.
\end{equation*}
If $\widehat q_\beta$ denotes the $\beta$ quantile of $\widehat L_{N,B}$, an equal-tailed $(1-\alpha)$ confidence interval is
\begin{equation*}
    \left[
    \widehat\tau_s-\frac{\widehat q_{1-\alpha/2}}{\sqrt N},
    \widehat\tau_s-\frac{\widehat q_{\alpha/2}}{\sqrt N}
    \right].
\end{equation*}
We require $b_N\rightarrow\infty$, $b_N/N\rightarrow0$, and
\begin{equation*}
    \sqrt{b_N}\left(b_{1,N}/b_N-N_1/N\right)\plim0.
\end{equation*}
A sufficient additional condition for fixed allocations is
\begin{equation}\label{eq:subsampling group moments}
    \condexpec{e_{i,s,r}^2}{D_i=d}=m_{s,d},
    \qquad r\in\T_s^{1,*},\quad d\in\{0,1\},
\end{equation}
where $m_{s,d}$ does not depend on $r$. Treatment-share fluctuations then add a common term to the surviving criteria and cancel from their softmax weights. Without this condition, fixing the group counts can omit relevant fluctuations in SSR contrasts. The condition holds automatically for a unique population minimizer and in the baseline simulations with group-independent errors and constant treatment effects. Alternatively, sample $b_N$ units uniformly without replacement from the full panel, allowing the treated count to vary; this preserves treatment-share fluctuations and does not require \eqref{eq:subsampling group moments}.

\begin{theorem}[Validity of subsampling]\label{theorem:subsampling}
    Suppose the conditions of Theorem \ref{theorem: unconditional asymptotic distribution} hold. Let $F_s$ be the distribution function of $L_s\equiv\bm w_s^{*\prime}\bm\eta_s$. If $b_N\rightarrow\infty$, $b_N/N\rightarrow0$, and $B\rightarrow\infty$, uniform unit subsampling satisfies $\widehat L_{N,B}(x)\plim F_s(x)$ at every continuity point of $F_s$. The same conclusion holds for the fixed allocations above under \eqref{eq:subsampling group moments}. If $F_s$ is continuous and strictly increasing at its $\alpha/2$ and $1-\alpha/2$ quantiles, the displayed interval has limiting coverage $1-\alpha$.
    \hfill $\blacksquare$
\end{theorem}

In the simulations and applications, we set $b_N=\lfloor N^{1/3}\rfloor$ and choose $b_{1,N}$ as the nearest feasible integer to $b_NN_1/N$, with $b_{0,N}=b_N-b_{1,N}$. This rounding satisfies the allocation-rate condition asymptotically. The Monte Carlo approximation requires $B\rightarrow\infty$. Coverage statements are pointwise, not uniform over shrinking criterion gaps. This limitation holds for any post-selection estimator \citep{Leeb_Potscher_2008}.

\subsection{Simulated Inference}\label{section:simulated inference}

Simulation-based inference for model-averaged estimators is studied in \citet{Lu_2015}, \citet{Zhang_Liu_2019_inference}, and \citet{Wang_et_al_2019}. In our setting, inference must retain the first-order criterion fluctuations among population SSR minimizers while discarding the other candidates. We therefore use a selector with a slower rate than the $\sqrt N$ criterion. This selector is used only for inference, not for computing the MADID point estimate. Define
\begin{equation*}
    \widehat{s}_{s,t}(a_N) = \exp\left( - a_N (\log(\widehat{\SSR}_{s,t} / N) - \log(\widehat{\rho}_s) )  \right),
\end{equation*}
where $\widehat{\rho}_s = \min\{\widehat{\SSR}_{s,1},...,\widehat{\SSR}_{s,T_0}\} / N$. If $a_N\rightarrow\infty$ and $a_N/\sqrt N\rightarrow0$, the selector tends to one for population SSR minimizers and to zero for the remaining candidates. 

\begin{lemma}\label{lemma:selector}
    Suppose that $\T_s^{1,*}$ is non-empty and Assumptions \ref{asm: unconditional setting} and \ref{asm: unconditional trends} hold. For any sequence $a_N$ such that $a_N \rightarrow \infty$ and $a_N / \sqrt{N} \rightarrow 0$, 
    \begin{equation*}
        \widehat{s}_{s,r}(a_N) \plim \mathds{1}(r \in \T_{s}^{1,*}) .
    \end{equation*}

    \hfill $\blacksquare$
    
\end{lemma}

We then simulate the joint centered DID and residual-square fluctuations for all candidates. The selectors make candidates outside $\T_s^{1,*}$ asymptotically negligible. For all $s \in \mathcal{S}$, let 
\begin{equation*}
    \widehat{\bm \psi}_{i,s} \equiv \left( \frac{D_i}{N_1 / N} - \frac{(1 - D_i)}{N_0 / N} \right) \left( \bm W_s \bm y_i - \widehat{\bm \gamma}_s - D_i \widehat{\bm \tau}_s \right),
\end{equation*}
and further define
\begin{equation*}
    \widehat{e}_{i,s,r} \equiv y_{i,s} - y_{i,r} - \widehat{\gamma}_{s,r} - D_i \widehat{\tau}_{s,r},
    \qquad
    \widehat{\xi}_{i,s,r} = \frac{\widehat{e}_{i,s,r}^2 - \widehat{\SSR}_{s,r} / N}{\widehat{\SSR}_{s,r} / N},
    \qquad 
    r \in \{1,...,T_0\}.
\end{equation*}
Stack these terms as $\widehat{\bm \xi}_{i,s} \equiv (\widehat{\xi}_{i,s,1},...,\widehat{\xi}_{i,s,T_0})'$, then let
\begin{equation*}
    \widehat{\bm \psi}_i \equiv (\widehat{\bm \psi}_{i,s}')_{s \in \mathcal{S}}',
    \qquad
    \widehat{\bm \xi}_i \equiv (\widehat{\bm \xi}_{i,s}')_{s \in \mathcal{S}}', 
    \qquad 
    \widehat{\bm V}_i \equiv
    \begin{pmatrix}
        \widehat{\bm \psi}_i\\
        \widehat{\bm \xi}_i
    \end{pmatrix}.
\end{equation*}
The overall asymptotic variance estimator is then 
\begin{equation*}
    \widehat{\bm \Sigma} \equiv \frac{1}{N} \isum \widehat{\bm V}_i \widehat{\bm V}_i'.
\end{equation*}
Note that $\widehat{\bm \Sigma}$ is \emph{not} an estimator of $\bm \Sigma$, which is defined for Theorem \ref{theorem: unconditional asymptotic distribution} with respect to $\T_s^{1,*}$ for all $s \in \mathcal{S}$. Its submatrix retaining all DID components and only the criterion components in $\T_s^{1,*}$ consistently estimates $\bm\Sigma$. The selectors remove the other criterion components asymptotically, not exactly in finite samples. 

To obtain simulated confidence intervals, we first obtain $B$ independent draws from 
\begin{equation*}
    \begin{pmatrix}
        \bm \eta^{(b)}\\
        \bm Z^{(b)}
    \end{pmatrix}
    \sim N \left( \bm 0, \widehat{\bm \Sigma} \right), 
    \qquad b = 1,...,B.
\end{equation*}
$\bm \eta^{(b)}$ can be used exactly as drawn. For $s \in \mathcal{S}$, let 
\begin{gather*}
    \widehat{\bm w}_s^{(b)} =
    \left( \sum_{t = 1}^{T_0} \widehat{s}_{s,t}(a_N) \exp \left( -Z_{s,t}^{(b)} / 2 \right) \right)^{-1}
    \begin{pmatrix}
        \widehat{s}_{s,1}(a_N) \exp\left( -Z_{s,1}^{(b)} / 2 \right)\\
        \vdots\\
        \widehat{s}_{s,T_0}(a_N) \exp\left( -Z_{s,T_0}^{(b)} / 2 \right)
    \end{pmatrix},
\end{gather*}
where $Z_{s,t}^{(b)}$ corresponds to the index $(s - T_0 - 1) T_0 + t$ in the vector $\bm Z^{(b)}$. The weights will always be well-defined because $\widehat{s}_{s,t}(a_N) = 1$ for $t$ corresponding to the period where $\widehat{\rho}_s = \widehat{\SSR}_{s,t} / N$. The simulated draw of the centered limiting root is 
\begin{equation*}
    \widehat{\tau}_s^{(b)} = \widehat{\bm w}_s^{(b)\prime} \bm \eta_s^{(b)},
\end{equation*}
where $\bm \eta_s^{(b)}$ is the $(s-T_0)$'th $T_0 \times 1$ subvector of $\bm \eta^{(b)}$. An equal-tailed $(1 - \alpha)$ confidence interval is 
\begin{equation*}
    \left[ \widehat{\tau}_s(\widehat{\bm w}_s) - \frac{\widehat{q}_{1 - \alpha / 2}}{\sqrt{N}}, \widehat{\tau}_s(\widehat{\bm w}_s) - \frac{\widehat{q}_{\alpha/2}}{\sqrt{N}} \right],
    \qquad 
    s \in \mathcal{S},
\end{equation*}
where $\widehat{q}_\beta$ is the $\beta$ quantile of the empirical distribution of $(\widehat{\tau}_s^{(b)})_{b = 1}^B$ without recentering. These draws are on the limiting-root scale, not the original ATT scale.

\begin{theorem}[Validity of simulated inference]\label{theorem:simulated inference}
    Suppose the conditions of Theorem \ref{theorem: unconditional asymptotic distribution} hold, $a_N\rightarrow\infty$, $a_N/\sqrt N\rightarrow0$, and $B\rightarrow\infty$. Conditional on the observed panel, the joint simulated distribution of $(\widehat\tau_s^{(b)})_{s\in\mathcal S}$ converges weakly in probability to that of $(L_s)_{s\in\mathcal S}$. For each $s$, if $F_s$ is continuous and strictly increasing at the two relevant quantiles, the displayed interval has limiting coverage $1-\alpha$.
    \hfill $\blacksquare$
\end{theorem}

\section{Simulations}\label{sec: simulations}

We use Monte Carlo experiments to assess three features of MADID: its performance relative to pre-testing and alternative time-weighting procedures, its sensitivity to the signal-to-noise ratio, and its behavior when candidate criteria are difficult to distinguish. Each design has five pre-treatment periods and one post-treatment period ($T_0 = 5$ and $T = 6$):
\begin{align*}
    y_{i,t}(0) &= c_i + f_t \gamma_i + u_{i,t},\\
    y_{i,t} &= y_{i,t}(0) + D_i \mathds{1}(t = 6)\tau,
\end{align*}
where $\tau = 1$, $D_i \sim \operatorname{Bernoulli}(1/2)$, $c_i = 1 + \eta_i^c$, and $\gamma_i = D_i + 2\eta_i^\gamma$, with $\eta_i^c,\eta_i^\gamma \stackrel{\mathrm{i.i.d.}}{\sim} N(0,1)$. In the baseline design, $u_{i,t} \sim N(0,\sigma_u^2)$ independently across units and periods. A candidate period $m$ is valid if $f_m = f_6$.

We consider the four fixed factor paths summarized in Table~\ref{tab:simulation dgp}. The sinusoidal, manual, and constant designs contain valid comparison periods, whereas the linear trend design represents complete misspecification.

\begin{table}[htbp]
\centering
\caption{Baseline simulation designs}
\label{tab:simulation dgp}
\small
\resizebox{\textwidth}{!}{
\begin{tabular}{lp{5.1cm}p{5.0cm}l}
\toprule
Design & Generating formula & Fixed factor path & Valid periods \\
\midrule
Sinusoidal & $f_{s+1}=\sin(\pi s+0.35)$, $s=0,\ldots,5$ & \shortstack[l]{$(0.343,-0.343,0.343,$\\$-0.343,0.343,-0.343)'$} & 2, 4 \\
Linear trend & $f_t=1+0.2(t-1)$, $t=1,\ldots,6$ & $(1,1.2,1.4,1.6,1.8,2)'$ & None \\
Manual & $\bm f=(1,1,1.5,1,1.5,1)'$ & $(1,1,1.5,1,1.5,1)'$ & 1, 2, 4 \\
Constant & $f_t=1$, $t=1,\ldots,6$ & $(1,1,1,1,1,1)'$ & All \\
\bottomrule
\end{tabular}}
\end{table}

For each candidate period $m \in \{1,\ldots,5\}$, MADID combines the corresponding $2 \times 2$ DID estimators according to
\begin{equation}\label{eq: simulation MADID}
    \widehat{\tau}_{6}(\widehat{\bm w}_6) \equiv \sum_{t = 1}^{5}\widehat w_{6,t}\widehat\tau_{6,t},
    \qquad
    \widehat w_{6,t} \equiv
    \frac{\exp(-IC_{6,t}/2)}{\sum_{r = 1}^{5}\exp(-IC_{6,r}/2)},
    \qquad
    IC_{6,r} \equiv \sqrt{N}\log(\widehat{SSR}_{6,r}/N).
\end{equation}
We compare MADID with the five individual DID estimators, pooled DID, TW-DID \citep{Schenk_2025}, and three pre-test procedures. For each $t\in\{1,\ldots,4\}$, we test $H_{0,t}: E(y_{i,5}-y_{i,t}\mid D_i=1)=E(y_{i,5}-y_{i,t}\mid D_i=0)$ using the difference in group means divided by $\{s_{1,t}^2/N_1+s_{0,t}^2/N_0\}^{1/2}$, where $s_{d,t}^2$ is the sample variance of $y_{i,5}-y_{i,t}$ in group $d$. Tests are two-sided at the 5\% level. All-pass reports pooled DID only if all four tests fail to reject. Minimum-$|t|$ DID uses the candidate in $\{1,\ldots,4\}$ with the smallest absolute statistic. Passing-period pooled DID averages the candidates that pass and always includes period 5, so it reports even if no tested period passes. The reported confidence intervals for these procedures use conventional standard errors after selection and do not account for the selection step.

Unless otherwise stated, $N = 200$ and $\sigma_u = 1$. Each cell is based on 1,000 Monte Carlo replications. We report bias, Monte Carlo standard deviation (SD), root mean squared error (RMSE), and the coverage probability (CP) of nominal 95\% confidence intervals. For All-pass, these measures are conditional on reporting; its reporting rate (RR) is the fraction of replications in which all four tests pass. The other procedures report in every replication. For MADID, CP is computed from the subsampling-quantile intervals in Section~\ref{section:subsampling}, using 499 draws per replication. Additional results for sample size, within-unit dependence, and average candidate weights are reported in Appendix Tables~\ref{tab:simulation N}--\ref{tab:simulation weights}.

\subsection{Comparison with alternative estimators}

Table~\ref{tab:simulation methods} reports the benchmark comparison. MADID is nearly unbiased in the sinusoidal, manual, and constant-factor designs, with coverage probabilities of 0.964, 0.948, and 0.956, respectively. In the manual design, its RMSE is lower than that of any individual valid DID, illustrating the precision gain from averaging across credible comparisons. When every comparison is valid, the constant-factor design yields nearly uniform average weights and MADID performs similarly to pooled DID. In the linear-trend design, MADID remains biased because every candidate is invalid. This is an identification failure rather than an inferential failure: in this design, all candidate biases have the same sign, so convex averaging cannot eliminate them. We note that the SD of the TW-DID estimator is uniformly smaller than MADID in every design and only larger than pooled DID under the constant factor where the Gauss-Markov assumptions hold. This fact is unsurprising, as TW-DID was designed to obtain relatively efficient estimators compared to traditional pooling. However, it severely under-covers relative to MADID in designs where some or all of the pre-treatment DID estimators are inconsistent.

\begin{table}[H]
\centering
\caption{Comparison of estimators for $N=200$ and $\sigma_u=1$}
\label{tab:simulation methods}
\scriptsize
\setlength{\tabcolsep}{4.0pt}
\begin{tabular}{lrrrrrrrr}
\toprule
\multicolumn{9}{l}{\textit{Panel A}} \\
\addlinespace[2pt]
& \multicolumn{4}{c}{Sinusoidal} & \multicolumn{4}{c}{Linear trend} \\
\cmidrule(lr){2-5}\cmidrule(lr){6-9}
Method & Bias & SD & RMSE & CP & Bias & SD & RMSE & CP \\
\midrule
DID 1 & -0.696 & 0.277 & 0.749 & 0.294 & 1.003 & 0.361 & 1.066 & 0.190 \\
DID 2 & -0.002 & 0.191 & 0.191 & 0.964 & 0.792 & 0.315 & 0.852 & 0.262 \\
DID 3 & -0.681 & 0.278 & 0.736 & 0.328 & 0.589 & 0.271 & 0.648 & 0.407 \\
DID 4 & 0.004 & 0.195 & 0.195 & 0.950 & 0.396 & 0.231 & 0.458 & 0.590 \\
DID 5 & -0.683 & 0.276 & 0.737 & 0.309 & 0.191 & 0.210 & 0.284 & 0.854 \\
Pooled DID & -0.412 & 0.187 & 0.452 & 0.434 & 0.594 & 0.239 & 0.640 & 0.273 \\
All-pass & -0.289 & 0.164 & 0.332 & 0.713 & 0.378 & 0.186 & 0.421 & 0.643 \\
Minimum-$|t|$ DID & -0.664 & 0.292 & 0.725 & 0.323 & 0.394 & 0.229 & 0.456 & 0.645 \\
Passing-period pooled DID & -0.629 & 0.297 & 0.696 & 0.259 & 0.334 & 0.224 & 0.402 & 0.608 \\
TW-DID & -0.130 & 0.164 & 0.209 & 0.895 & 0.222 & 0.192 & 0.293 & 0.790 \\
MADID & -0.012 & 0.168 & 0.168 & 0.964 & 0.251 & 0.200 & 0.321 & 0.861 \\
\midrule
\multicolumn{9}{l}{\textit{Panel B}} \\
\addlinespace[2pt]
& \multicolumn{4}{c}{Manual} & \multicolumn{4}{c}{Constant} \\
\cmidrule(lr){2-5}\cmidrule(lr){6-9}
Method & Bias & SD & RMSE & CP & Bias & SD & RMSE & CP \\
\midrule
DID 1 & -0.005 & 0.205 & 0.205 & 0.943 & 0.002 & 0.204 & 0.204 & 0.941 \\
DID 2 & 0.006 & 0.206 & 0.206 & 0.940 & -0.002 & 0.201 & 0.201 & 0.947 \\
DID 3 & -0.504 & 0.240 & 0.558 & 0.449 & -0.005 & 0.199 & 0.199 & 0.954 \\
DID 4 & -0.003 & 0.202 & 0.202 & 0.944 & -0.006 & 0.202 & 0.202 & 0.943 \\
DID 5 & -0.506 & 0.243 & 0.561 & 0.466 & -0.005 & 0.196 & 0.196 & 0.949 \\
Pooled DID & -0.202 & 0.166 & 0.261 & 0.768 & -0.003 & 0.155 & 0.155 & 0.957 \\
All-pass & -0.161 & 0.168 & 0.233 & 0.837 & -0.005 & 0.153 & 0.153 & 0.961 \\
Minimum-$|t|$ DID & -0.437 & 0.264 & 0.510 & 0.544 & -0.005 & 0.182 & 0.181 & 0.975 \\
Passing-period pooled DID & -0.398 & 0.267 & 0.479 & 0.480 & -0.003 & 0.161 & 0.161 & 0.951 \\
TW-DID & -0.092 & 0.165 & 0.189 & 0.911 & -0.005 & 0.158 & 0.158 & 0.955 \\
MADID & -0.024 & 0.169 & 0.171 & 0.948 & -0.005 & 0.159 & 0.159 & 0.956 \\
\bottomrule
\end{tabular}
\par\smallskip
\begin{minipage}{0.98\textwidth}
\footnotesize
\textit{Notes:} The All-pass reporting rates (RR) are 0.181, 0.171, 0.227, and 0.842 in the sinusoidal, linear-trend, manual, and constant-factor designs, respectively.
\end{minipage}
\end{table}

The pre-test procedures perform poorly in designs containing invalid comparisons. A small pre-trend statistic identifies a period that resembles another pre-treatment period, not necessarily one that shares the post-treatment factor. Accordingly, minimum-$|t|$ and passing-period pooling can place substantial weight on invalid comparisons. All-pass is additionally selective. In the constant-factor design, by contrast, every candidate is valid, and the differences among aggregation rules primarily reflect sampling variation rather than specification bias. These results complement the findings of \citet{Roth_2022} regarding the problematic nature of common diagnostics. 

\subsection{Signal strength}

Table~\ref{tab:simulation sigma} varies the idiosyncratic error standard deviation. In the sinusoidal and manual designs, increasing $\sigma_u$ makes the candidate SSRs harder to distinguish, producing more diffuse weights and increasing both bias and RMSE. The constant-factor design remains essentially unbiased throughout because every comparison is valid. In the linear-trend design, reducing $\sigma_u$ makes MADID more closely select the last pre-treatment period but cannot remove its population bias.

\begin{table}[H]
\centering
\caption{MADID across error standard deviations ($N=200$)}
\label{tab:simulation sigma}
\small
\begin{tabular}{lrrrrr}
\toprule
Design & $\sigma_u$ & Bias & SD & RMSE & CP \\
\midrule
Sinusoidal & 0.25 & 0.001 & 0.043 & 0.043 & 0.974 \\
 & 0.50 & 0.002 & 0.087 & 0.087 & 0.971 \\
 & 1.00 & -0.012 & 0.168 & 0.168 & 0.964 \\
 & 2.00 & -0.163 & 0.346 & 0.382 & 0.930 \\
 & 3.00 & -0.311 & 0.496 & 0.585 & 0.922 \\
\midrule
Linear trend & 0.25 & 0.198 & 0.080 & 0.213 & 0.707 \\
 & 0.50 & 0.205 & 0.114 & 0.234 & 0.861 \\
 & 1.00 & 0.251 & 0.200 & 0.321 & 0.861 \\
 & 2.00 & 0.412 & 0.358 & 0.546 & 0.821 \\
 & 3.00 & 0.521 & 0.531 & 0.744 & 0.813 \\
\midrule
Manual & 0.25 & 0.001 & 0.042 & 0.042 & 0.968 \\
 & 0.50 & 0.000 & 0.084 & 0.084 & 0.958 \\
 & 1.00 & -0.024 & 0.169 & 0.171 & 0.948 \\
 & 2.00 & -0.143 & 0.320 & 0.350 & 0.939 \\
 & 3.00 & -0.178 & 0.484 & 0.516 & 0.929 \\
\midrule
Constant factor & 0.25 & -0.000 & 0.039 & 0.039 & 0.958 \\
 & 0.50 & 0.003 & 0.080 & 0.080 & 0.958 \\
 & 1.00 & -0.005 & 0.159 & 0.159 & 0.956 \\
 & 2.00 & -0.001 & 0.315 & 0.315 & 0.955 \\
 & 3.00 & 0.012 & 0.478 & 0.478 & 0.956 \\
\bottomrule
\end{tabular}
\end{table}

\subsection{Loading mean--variance trade-off}

The SSR criterion is informative because invalid comparisons generate additional dispersion through heterogeneous factor loadings. To separate this source of information from the bias induced by selection on the loadings, Table~\ref{tab:simulation gamma grid} varies the treated--control mean difference $\Delta_\gamma$ and the common conditional variance $V_\gamma$ independently. The experiment retains the manual factor path, $N = 200$, and $\sigma_u = 1$. A larger $\Delta_\gamma$ magnifies the bias from invalid comparisons, whereas a larger $V_\gamma$ makes those comparisons easier for the SSR criterion to detect.

\begin{table}[H]
\centering
\caption{Loading mean-gap and variance grid}
\label{tab:simulation gamma grid}
\small
\begin{tabular}{rrrrrrrr}
\toprule
& & \multicolumn{3}{c}{Pooled DID} & \multicolumn{3}{c}{MADID} \\
\cmidrule(lr){3-5}\cmidrule(lr){6-8}
$\Delta_\gamma$ & $V_\gamma$ & Bias & RMSE & CP & Bias & RMSE & CP \\
\midrule
0.5 & 0.25 & -0.101 & 0.187 & 0.900 & -0.091 & 0.186 & 0.922 \\
0.5 & 1.00 & -0.102 & 0.188 & 0.900 & -0.062 & 0.176 & 0.944 \\
0.5 & 4.00 & -0.102 & 0.195 & 0.906 & -0.014 & 0.169 & 0.950 \\
\midrule
1.0 & 0.25 & -0.201 & 0.255 & 0.738 & -0.180 & 0.248 & 0.806 \\
1.0 & 1.00 & -0.202 & 0.257 & 0.746 & -0.121 & 0.209 & 0.899 \\
1.0 & 4.00 & -0.202 & 0.261 & 0.765 & -0.024 & 0.171 & 0.948 \\
\midrule
3.0 & 0.25 & -0.601 & 0.621 & 0.033 & -0.535 & 0.588 & 0.268 \\
3.0 & 1.00 & -0.602 & 0.622 & 0.034 & -0.358 & 0.424 & 0.563 \\
3.0 & 4.00 & -0.602 & 0.625 & 0.058 & -0.067 & 0.186 & 0.951 \\
\bottomrule
\end{tabular}
\end{table}

The results expose the economically relevant limitation of the procedure. Holding $\Delta_\gamma = 1$, increasing $V_\gamma$ from 0.25 to 4 reduces MADID's RMSE from 0.248 to 0.171 and raises coverage from 0.806 to 0.948. When the mean difference is large but loading heterogeneity is weak, however, invalid comparisons can be substantially biased without producing a sufficiently large variance signal. In the most adverse cell, $(\Delta_\gamma,V_\gamma)=(3,0.25)$, MADID has bias $-0.535$ and coverage 0.268. Thus, the simulations reinforce the role of Assumption~\ref{asm: unconditional trends}: MADID performs best when departures from parallel trends are accompanied by enough cross-sectional heterogeneity to be visible in the candidate SSRs.

\subsection{Near-tie designs}

The final experiment varies $\delta\in\{0.05,0.10,0.25\}$ from $\bm f=(1,1.5,1.5,1,1,1.5)'$ and $Var(u_{i,t})=1$. The factor-only design sets $f_3=f_5=1.5+\delta$; the factor--variance design sets $f_3=1.5+\delta$ and $Var(u_{i,3})=1-\delta$; the variance-only design changes only $Var(u_{i,3})$ to $1-\delta$. With $V_\gamma=4$ and independent errors, Table~\ref{tab:simulation near tie} reports $g_\rho$, the population SSR difference from valid period 2: $4\delta^2$ for each perturbed candidate in the factor-only design, $4\delta^2-\delta$ for period 3 in the factor--variance design, and $-\delta$ for period 3 in the variance-only design. For $0<\delta<0.25$, the invalid period 3 has lower population SSR in the factor--variance design; at $\delta=0.25$, it ties with valid candidates despite having a different DID estimand. The favored candidate in the variance-only design remains valid.

\begin{table}[H]
\centering
\caption{MADID under near-tie designs ($\sigma_u=1$)}
\label{tab:simulation near tie}
\scriptsize
\renewcommand{\arraystretch}{0.88}
\begin{tabular}{lrrrrrrr}
\toprule
Perturbation & $\delta$ & $g_\rho$ & $N$ & Bias & SD & RMSE & CP \\
\midrule
Factor + variance & 0.050 & -0.040 & 50 & 0.120 & 0.336 & 0.356 & 0.963 \\
 & 0.050 &  & 200 & 0.024 & 0.176 & 0.178 & 0.950 \\
 & 0.050 &  & 1000 & -0.029 & 0.080 & 0.085 & 0.890 \\
 & 0.100 & -0.060 & 50 & 0.105 & 0.348 & 0.363 & 0.965 \\
 & 0.100 &  & 200 & -0.013 & 0.172 & 0.172 & 0.935 \\
 & 0.100 &  & 1000 & -0.060 & 0.078 & 0.098 & 0.795 \\
 & 0.250 & 0.000 & 50 & 0.049 & 0.354 & 0.358 & 0.959 \\
 & 0.250 &  & 200 & -0.063 & 0.187 & 0.197 & 0.886 \\
 & 0.250 &  & 1000 & -0.127 & 0.090 & 0.155 & 0.503 \\
\midrule
Factor only & 0.050 & 0.010 & 50 & 0.038 & 0.340 & 0.342 & 0.964 \\
 & 0.050 &  & 200 & -0.008 & 0.166 & 0.166 & 0.943 \\
 & 0.050 &  & 1000 & -0.033 & 0.074 & 0.081 & 0.916 \\
 & 0.100 & 0.040 & 50 & 0.016 & 0.325 & 0.325 & 0.966 \\
 & 0.100 &  & 200 & -0.032 & 0.167 & 0.170 & 0.937 \\
 & 0.100 &  & 1000 & -0.056 & 0.079 & 0.097 & 0.816 \\
 & 0.250 & 0.250 & 50 & -0.014 & 0.340 & 0.340 & 0.967 \\
 & 0.250 &  & 200 & -0.072 & 0.175 & 0.189 & 0.896 \\
 & 0.250 &  & 1000 & -0.065 & 0.086 & 0.108 & 0.795 \\
\midrule
Variance only & 0.050 & -0.050 & 50 & 0.142 & 0.353 & 0.380 & 0.957 \\
 & 0.050 &  & 200 & 0.048 & 0.177 & 0.183 & 0.956 \\
 & 0.050 &  & 1000 & 0.003 & 0.079 & 0.079 & 0.929 \\
 & 0.100 & -0.100 & 50 & 0.129 & 0.332 & 0.356 & 0.978 \\
 & 0.100 &  & 200 & 0.040 & 0.180 & 0.184 & 0.940 \\
 & 0.100 &  & 1000 & 0.004 & 0.077 & 0.077 & 0.945 \\
 & 0.250 & -0.250 & 50 & 0.120 & 0.358 & 0.378 & 0.954 \\
 & 0.250 &  & 200 & 0.023 & 0.173 & 0.175 & 0.948 \\
 & 0.250 &  & 1000 & 0.002 & 0.080 & 0.080 & 0.929 \\
\bottomrule
\end{tabular}
\end{table}

Weak separation matters primarily when a nearly tied candidate is invalid. In the factor-only design, each fixed $\delta>0$ preserves population separation, but the finite-sample criterion can remain noisy when $4\delta^2$ is small. In the factor--variance design, the identification condition fails: an invalid candidate minimizes or ties for the minimum population SSR, so the ATT-centered limit in Theorem~\ref{theorem: unconditional asymptotic distribution} does not apply. With $N=1000$, coverage falls from 0.890 at $\delta=0.05$ to 0.503 at $\delta=0.25$. By contrast, coverage is more stable in the variance-only design because the favored comparison remains valid. Thus, random weights among valid minimizers are distinct from weight on an invalid minimizer; the latter can generate persistent specification bias.

\section{Applications}

We consider two applications with markedly different cross-sectional sample sizes and pre-treatment trends. The first examines the effect of the federal Opportunity Zone program on census tract-level employment and unemployment using 29,072 tracts. The event-study estimates display pronounced pre-trends, and the large sample sharply separates the candidate criteria: MADID places essentially all weight on the last pre-treatment period. The resulting estimates are about 79\% and 75\% smaller in magnitude than pooled DID for employment and unemployment, respectively. The second application studies the effect of the US Supreme Court decision in \textit{Dobbs v. Jackson Women's Health Organization} using 37 jurisdictions. The event-study estimates here provide little evidence of pre-trends. MADID again favors the nearest pre-treatment period, but the smaller sample leaves non-negligible weight on earlier comparisons. These applications illustrate how the same criterion can produce concentrated or diffuse weights depending on both criterion separation and sampling uncertainty.

\subsection{Opportunity Zones}

The 2017 Tax Cuts and Jobs Act significantly overhauled the US tax system.\footnote{See \citet{Gale_et_al_2024} for an overview of the legislation and an analysis of its economic impacts.} Although its main purpose was to reform the individual and business tax codes, it also created the `Opportunity Zone' census-tract designation. The program provided tax incentives for investments in businesses or real estate located in Opportunity Zones, either directly or through Qualified Opportunity Funds that held at least 90\% of their assets in designated zones.

There is an extensive literature on `place-based' policies, or policies that target specific geographic areas. Economic theory suggests that such interventions may generate long-run benefits through agglomeration economies, although the empirical evidence is mixed. \citet{Ham_et_al_2011} and \citet{Busso_Gregory_Kline_2013} find significant positive employment effects of Empowerment Zones, another federal place-based policy.\footnote{See \citet{Neumark_Simpson_2015} for a review of the Empowerment Zone program.} \citet{Neumark_Young_2019} reexamine these estimates and argue that earlier work did not adequately address pre-trend violations. We use American Community Survey (ACS) five-year estimates of census tract-level labor-market outcomes from 2013--2019, matching the period studied by \citet{Freedman_Khanna_Neumark_2023}. The samples differ because they use restricted-access annual ACS data, whereas we use publicly available five-year estimates. Their inverse-propensity-score-weighted analysis finds no employment effect, a small positive earnings effect, and a small reduction in poverty; their event-study estimates also show clear evidence of non-parallel trends. Because our five-year ACS outcome windows overlap, the coefficients describe changes in rolling multi-year outcomes rather than annual policy effects.

State governors had the ability to designate tracts as Opportunity Zones if they qualified as a ``low-income community" (LIC) or were contiguous with a LIC. An LIC designation required ``a poverty rate of 20\% or median family income less than or equal to 80\% of the greater metropolitan area or statewide median family income (just statewide for rural tracts)" \citep{Freedman_Khanna_Neumark_2023}. We restrict our sample to tracts with at least a 20\% poverty rate or less than or equal to 80\% of statewide median income in 2017, as all designations were made by 2018. We perform our analysis on tract-level employment and unemployment rates. We only include tracts that have a population of at least 1,000 in every year. Our final data set contains $N = 29,072$ total census tracts, with $N_1 = 6,759$ receiving Opportunity Zone designation. 

\begin{figure}[h]
    \centering
    \caption{Effect of base period on event study coefficients}

    \label{fig: event study base periods}
    \includegraphics[width=\linewidth]{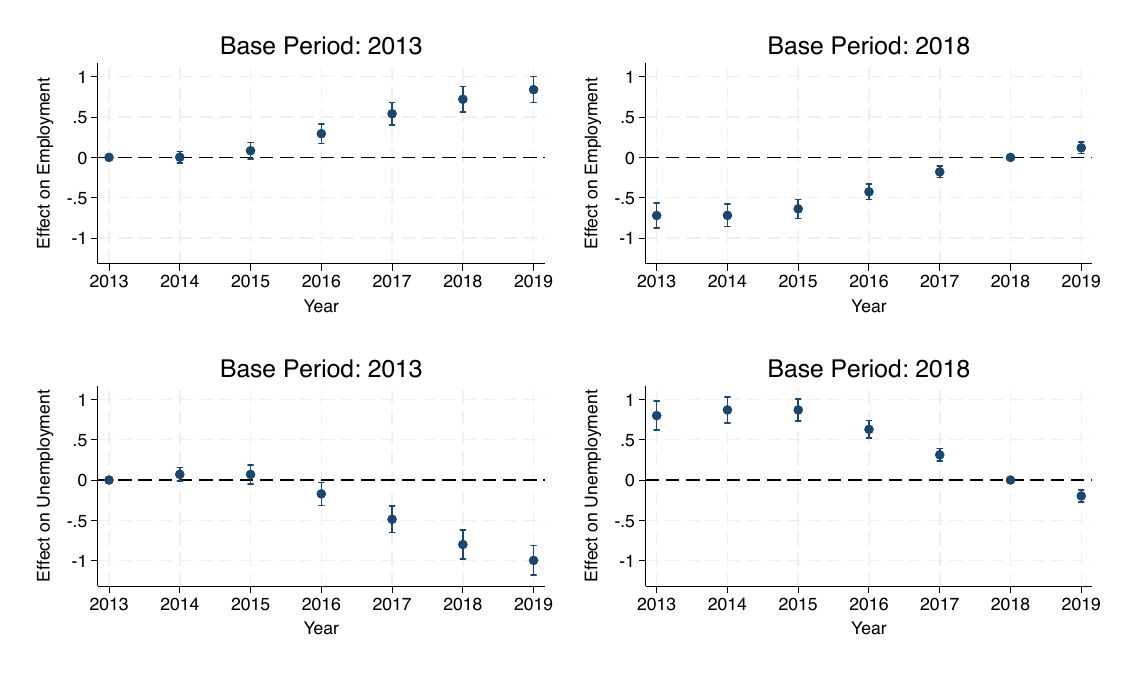} 

    \begin{flushleft}
    \footnotesize 
    \textit{Notes:} Event study plots for the effect of Opportunity Zone designation on employment and unemployment rates. Vertical bars represent 95\% confidence intervals based on standard errors clustered at the census tract level. 
    \end{flushleft}
    
\end{figure}

Figure \ref{fig: event study base periods} provides two event studies for the effect of Opportunity Zone designation on employment and unemployment rates, respectively. Using a 2013 base period, the estimated ATTs on employment and unemployment are 0.84 and -1.00 percentage points, respectively. If we instead use a 2018 base period, the respective estimated ATTs are 0.12 and -0.20 percentage points. These results show that the estimated effects are sensitive to the base period, but the event-study plots do not identify a valid comparison. A transitory, mean-reverting selection shock could generate an ``Ashenfelter's dip'', in which case the nearest pre-treatment period may be the most contaminated. Alternatively, under a scalar factor model with a stable treated--control loading mean difference, smaller factor gaps imply smaller absolute DID bias. MADID may favor a recent base period if its residual variance is also smaller, as in our `Linear trend' Monte Carlo design. However, the SSR criterion ranks residual dispersion rather than causal bias, and time proximity alone does not establish that these rankings coincide. We therefore interpret the concentration of weight on 2018 as a feature of the estimator in this sample, not evidence that this comparison satisfies parallel trends.

Table \ref{table: oz} reports DID and MADID estimates of the employment and unemployment effects of Opportunity Zone designation. The estimated magnitudes decline sharply as the base period approaches treatment: the 2013-based employment and unemployment estimates are, respectively, about seven and five times as large in magnitude as their 2018-based counterparts. In both cases, MADID places essentially all weight on 2018. We construct the MADID subsampled 95\% confidence intervals using a subsample size of $b=\lfloor N^{1/3}\rfloor=30$, allocating $b_1=7$ treated units and $b_0=23$ control units in each sample. Despite the large sample size and near equality of the point estimates, the MADID confidence interval for the employment effect is slightly wider compared to that for the 2018 DID estimator. This discrepancy shows that nearly equal point estimates need not yield identical intervals under different inference procedures. An equally weighted average of the 2013--2015 DID estimates would instead be about 6.8 times as large as MADID for employment and more than five times as large in magnitude for unemployment. 

\begin{table}[htbp]
  \centering
  \caption{Estimated effect of Opportunity Zone designation}

  \label{table: oz}
  
    \begin{tabular}{lcccc}
    \toprule
    \multicolumn{5}{c}{\textbf{Employment}} \\
    \midrule
          & \multicolumn{1}{l}{\textbf{Estimate}} & \multicolumn{1}{l}{\textbf{CI lower bound}} & \multicolumn{1}{l}{\textbf{CI upper bound}} & \multicolumn{1}{l}{\textbf{MADID weight}} \\
    \midrule
    \textbf{Pooled} & 0.5660 & 0.4574 & 0.6746 & - \\
    \boldmath{}\textbf{$\text{DID}_{2013}$}\unboldmath{} & 0.8391 & 0.6770 & 1.0012 & 0.0000 \\
    \boldmath{}\textbf{$\text{DID}_{2014}$}\unboldmath{} & 0.8371 & 0.6812 & 0.9930 & 0.0000 \\
    \boldmath{}\textbf{$\text{DID}_{2015}$}\unboldmath{} & 0.7566 & 0.6187 & 0.8945 & 0.0000 \\
    \boldmath{}\textbf{$\text{DID}_{2016}$}\unboldmath{} & 0.5457 & 0.4262 & 0.6653 & 0.0000 \\
    \boldmath{}\textbf{$\text{DID}_{2017}$}\unboldmath{} & 0.2982 & 0.1997 & 0.3968 & 0.0000 \\
    \boldmath{}\textbf{$\text{DID}_{2018}$}\unboldmath{} & 0.1192 & 0.0470 & 0.1913 & 1.0000 \\
    \textbf{MADID} & 0.1192 & 0.0412 & 0.1921 & - \\
    \midrule
    \multicolumn{5}{c}{\textbf{Unemployment}} \\
    \midrule
          & \multicolumn{1}{l}{\textbf{Estimate}} & \multicolumn{1}{l}{\textbf{CI lower bound}} & \multicolumn{1}{l}{\textbf{CI upper bound}} & \multicolumn{1}{l}{\textbf{MADID weight}} \\
    \midrule
    \textbf{Pooled} & -0.7762 & -0.8944 & -0.6579 & - \\
    \boldmath{}\textbf{$\text{DID}_{2013}$}\unboldmath{} & -0.9955 & -1.1772 & -0.8139 & 0.0000 \\
    \boldmath{}\textbf{$\text{DID}_{2014}$}\unboldmath{} & -1.0658 & -1.2415 & -0.8900 & 0.0000 \\
    \boldmath{}\textbf{$\text{DID}_{2015}$}\unboldmath{} & -1.0654 & -1.2198 & -0.9111 & 0.0000 \\
    \boldmath{}\textbf{$\text{DID}_{2016}$}\unboldmath{} & -0.8240 & -0.9551 & -0.6929 & 0.0000 \\
    \boldmath{}\textbf{$\text{DID}_{2017}$}\unboldmath{} & -0.5090 & -0.6139 & -0.4041 & 0.0000 \\
    \boldmath{}\textbf{$\text{DID}_{2018}$}\unboldmath{} & -0.1971 & -0.2703 & -0.1240 & 1.0000 \\
    \textbf{MADID} & -0.1971 & -0.2673 & -0.1148 & - \\
    \bottomrule
    \end{tabular}%
    \begin{flushleft}
     \footnotesize
     \textit{Notes:} Estimates of the ATT of Opportunity Zone designation on employment and unemployment. Results are reported for the DID estimator that pools over all pre-treatment periods (`Pooled'), the individual $2 \times 2$ DID estimators, and the MADID estimator. $\text{DID}_{t}$ corresponds to the DID estimator between periods $t$ and $2019$ for $t \in \{2013,...,2018\}$. DID 95\% confidence intervals are constructed using standard errors clustered at the census tract level. MADID 95\% confidence intervals are obtained using 4,999 subsample draws, where $b = 30$ units are sampled without replacement ($b_1 = 7$, $b_0 = 23$), per \citet{Politis_Romano_2010}. The `MADID weight' column corresponds to the estimated weights in equation \eqref{eq: MADID weight}.
    \end{flushleft}
  
\end{table}%

\subsection{\textit{Dobbs v. Jackson Women's Health Organization}}

In 2022, the United States Supreme Court held that the Constitution does not confer a right to abortion, allowing states to revise their abortion laws. Using state-level data, \citet{Dench_et_al_2024} provide the first study of the effects of the \textit{Dobbs} decision on natality by studying the impact on log of births and fertility, where fertility is the number of births divided by the total female population and multiplied by 1000. Treatment indicates whether a state had effectively banned abortion by the end of 2022. We treat 2023 as the first post-treatment year because births affected by the new laws would begin to occur in early 2023. The original analysis uses only the first six months of each year because of data availability; we extend it to annual data while retaining the same sample of $N=37$ jurisdictions, including the District of Columbia, of which $N_1=12$ enacted abortion bans.\footnote{Some states were excluded because of legal uncertainty. For example, Florida enacted a 15-week ban in July 2022, but legal challenges affected its status during the sample period.}

\citet{Dench_et_al_2024} use the synthetic difference-in-differences estimator of \citet{Arkhangelsky_et_al_2021}. Its unit- and time-weighting scheme accommodates many empirically relevant departures from parallel trends. Its asymptotic guarantees, however, require restrictions on both panel dimensions. Assumption 2(i) of \citet{Arkhangelsky_et_al_2021} requires $T_0\rightarrow\infty$ and $\max\{N_1,T-T_0\}\rightarrow\infty$. The 2019--2023 sample used by \citet{Dench_et_al_2024} contains only four pre-treatment periods.

\begin{figure}[h]
\caption{Event study estimates}

\label{fig: dobbs event studies}

    \begin{subfigure}{0.48\textwidth}
        \includegraphics[width=\linewidth]{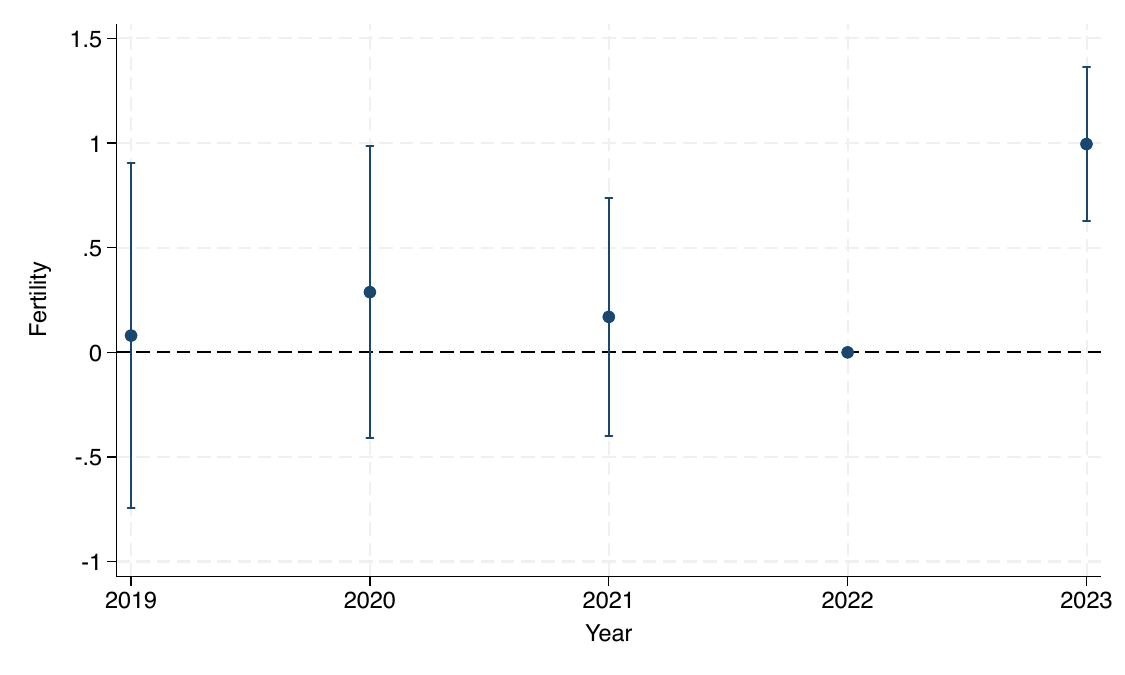}
        \caption{Fertility}
    \end{subfigure}
    \hfill
    \begin{subfigure}{0.48\textwidth}
        \includegraphics[width=\linewidth]{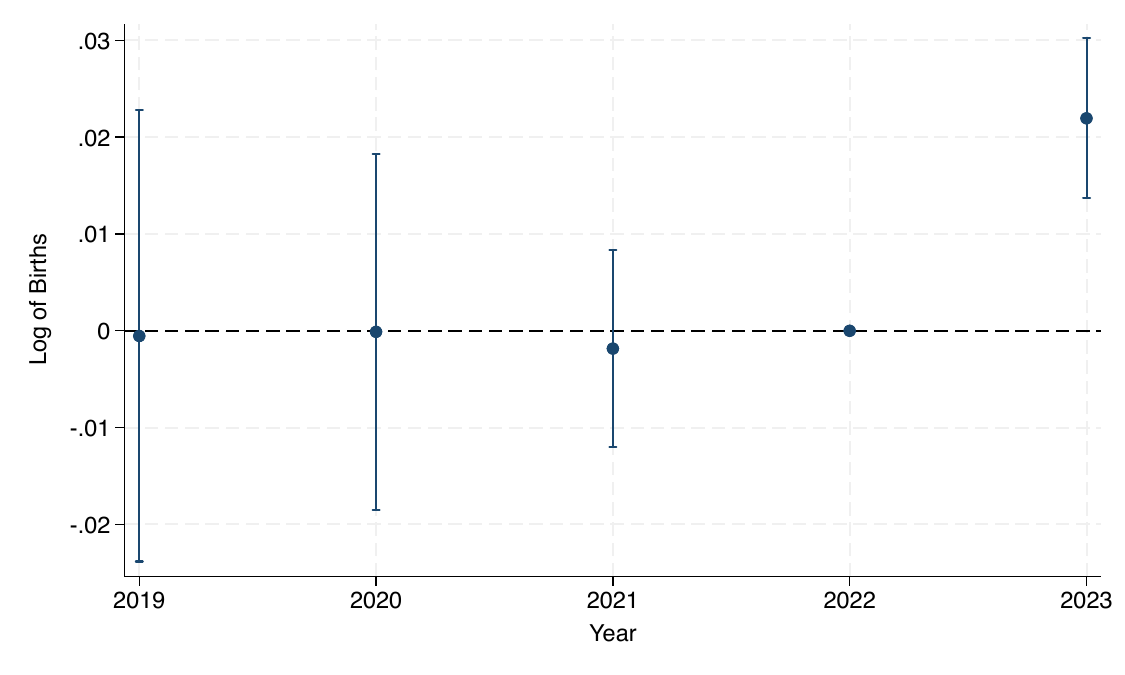}
        \caption{Log of births}
    \end{subfigure}

    \begin{flushleft}
    \footnotesize 
    \textit{Notes:} Event study plots for the effect of post-\textit{Dobbs} bans on fertility and log of births. Fertility is defined as 1,000 times the number of total births divided by the number of women. Vertical bars represent 95\% confidence intervals clustered at the state level. 
    \end{flushleft}
\end{figure}

The decision was seen as politically unpopular at the time \citep{YouGov}, but its popularity does not establish parallel trends. Figure \ref{fig: dobbs event studies} plots event study coefficients for fertility and log of births in banned states. The estimates show no statistically significant pre-trend; with this short panel and small sample, however, a failure to reject cannot establish the identifying assumption. 

Table \ref{table: dobbs} includes MADID estimates, along with pooled and the individual $2\times 2$ DID estimates. The MADID weights are mostly dominated by the 2022 base period, but the 2021 base has a 9.1\% weight when estimating fertility and 4.54\% weight for log of births. It is interesting to note that the 2.2\% increase in births is almost identical between MADID and the pooled estimator, which itself is close to the 2.3\% increase estimated by \citet{Dench_et_al_2024}. However, the MADID estimator is about 13\% larger than the pooled estimator for fertility change. Its confidence intervals are narrower than the pooled intervals but wider than those of the 2022-based DID estimator. Even if all periods satisfy parallel trends, equal time weights need not be efficient under unequal variances or serial covariance. With only $b=3$ observations per subsample, the asymptotic coverage result provides no finite-sample guarantee.

\begin{table}[htbp]
  \centering
  \caption{Estimated effect of abortion bans in 2023}
  \label{table: dobbs}
    \begin{tabular}{lcccc}
    \multicolumn{5}{c}{\textbf{Log of births}} \\
    \midrule
          & \multicolumn{1}{l}{\textbf{Estimate}} & \multicolumn{1}{l}{\textbf{CI lower bound}} & \multicolumn{1}{l}{\textbf{CI upper bound}} & \multicolumn{1}{l}{\textbf{MADID weight}} \\
    \midrule
    \textbf{Pooled} & 0.0226 & 0.0057 & 0.0395 & - \\
    \boldmath{}\textbf{$\text{DID}_{2019}$}\unboldmath{} & 0.0225 & -0.0053 & 0.0502 & 0.0005 \\
    \boldmath{}\textbf{$\text{DID}_{2020}$}\unboldmath{} & 0.0221 & -0.0009 & 0.0450 & 0.0018 \\
    \boldmath{}\textbf{$\text{DID}_{2021}$}\unboldmath{} & 0.0238 & 0.0106 & 0.0370 & 0.0454 \\
    \boldmath{}\textbf{$\text{DID}_{2022}$}\unboldmath{} & 0.0219 & 0.0138 & 0.0301 & 0.9524 \\
    \textbf{MADID} & 0.0220 & 0.0085 & 0.0363 & - \\
    \midrule
    \multicolumn{5}{c}{\textbf{Fertility}} \\
    \midrule
          & \multicolumn{1}{l}{\textbf{Estimate}} & \multicolumn{1}{l}{\textbf{CI lower bound}} & \multicolumn{1}{l}{\textbf{CI upper bound}} & \multicolumn{1}{l}{\textbf{MADID weight}} \\
    \midrule
    \textbf{Pooled} & 0.8612 & 0.3101 & 1.4124 & - \\
    \boldmath{}\textbf{$\text{DID}_{2019}$}\unboldmath{} & 0.9154 & -0.0402 & 1.8710 & 0.0025 \\
    \boldmath{}\textbf{$\text{DID}_{2020}$}\unboldmath{} & 0.7079 & -0.0956 & 1.5115 & 0.0042 \\
    \boldmath{}\textbf{$\text{DID}_{2021}$}\unboldmath{} & 0.8260 & 0.2789 & 1.3732 & 0.0910 \\
    \boldmath{}\textbf{$\text{DID}_{2022}$}\unboldmath{} & 0.9956 & 0.6311 & 1.3601 & 0.9024 \\
    \textbf{MADID} & 0.9787 & 0.5246 & 1.4794 & - \\
    \bottomrule
    \end{tabular}%
  \begin{flushleft}
     \footnotesize
     \textit{Notes:} Estimates of the ATT of the \textit{Dobbs} decision on log of births and fertility. Results are reported for the DID estimator that pools over all pre-treatment periods (`Pooled'), the individual $2 \times 2$ DID estimators, and the MADID estimator. $\text{DID}_{t}$ corresponds to the DID estimator between periods $t$ and $2023$ for $t \in \{2019,...,2022\}$. DID 95\% confidence intervals are constructed using standard errors clustered at the state level. MADID 95\% confidence intervals are obtained using 4,999 subsample draws, where $b = 3$ units are sampled without replacement ($b_1 = 1$, $b_0 = 2$), per \citet{Politis_Romano_2010}. The `MADID weight' column corresponds to the estimated weights in equation \eqref{eq: MADID weight}.
    \end{flushleft}
\end{table}%

\section{Conclusion}

We propose the model averaged difference-in-differences (MADID) estimator as an alternative to specification choice based on pre-trend tests. Under our maintained conditions, MADID is consistent when at least one pre-treatment period satisfies parallel trends and the population SSR criterion separates valid from invalid comparisons. Because its data-dependent weights can remain random asymptotically, the joint limiting $\sqrt{N}$-distribution of the post-treatment ATT estimators is generally non-Gaussian. We therefore develop $K$-sample subsampling and simulation-based inference. 

Our first empirical application examines the effect of place-based policies on local labor market outcomes via `Opportunity Zone' tax schemes. Preliminary event studies provide strong evidence of parallel trend violations. The MADID estimator places almost all weight on the last pre-treatment period. Confidence intervals obtained by subsampling are similar to the cluster-robust confidence interval on the last-period DID estimator. The second empirical application investigates the impact of the criminalization of abortion on fertility. Unlike the Opportunity Zone application, we have a relatively small sample of states with no clear pre-trend violations. The MADID estimator smooths more evenly over all pre-periods, though the last period still receives at least 90\% weight in both outcome specifications. The MADID intervals are shorter than the pooled intervals but wider than those of the last-period DID estimator. Their finite-sample coverage remains uncertain with $b=3$ subsampling. 

Extensions of the basic $2\times2$ DID estimator to MADID require corresponding identification conditions and influence-function arguments. Most empirical applications use covariates to justify a conditional parallel trends assumption. A model averaging approach would need to allow both for uncertainty between time periods and uncertainty between models within time periods. We should also consider nonlinear models and doubly robust methods that model both the outcome and the conditional probability of receiving treatment. Again, the problem is non-trivial because weights would need to balance both averaging problems simultaneously. Finally, we may consider alternative scales to the MADID weights to obtain efficiency gains and improved finite-sample performance. The factor $1/2$ follows the usual information-criterion convention, $\text{IC}=-2\log L+\text{penalty}$, with weights proportional to $\exp(-\text{IC}/2)$ \citep{Buckland_et_al_1997}. \citet{Athey_et_al_2026} estimate a tuning parameter associated with their distance weights. However, their distance measures are deterministic and not themselves estimated from the data like our likelihood functions.

\newpage~\bibliography{references.bib}

@article{Brown_Butts_2025,
  title={Dynamic treatment effect estimation with interactive fixed effects and short panels},
  author={Brown, Nicholas L and Butts, Kyle},
  journal={Journal of Econometrics},
  volume={250},
  pages={106013},
  year={2025},
  publisher={Elsevier}
}

@article{Brown_et_al_2023,
  title={Direct and Indirect Treatment Effects With Time-Varying Covariates},
  author={Brown, Nicholas L and Butts, Kyle and Westerlund, Joakim},
  journal={Journal of Applied Econometrics},
  year={2026},
  publisher={Wiley Online Library}
}

@article{Callaway_SantAnna_2021,
  title    = {Difference-in-Differences with Multiple Time Periods},
  journal  = {Journal of Econometrics},
  author   = {Callaway, Brantly and Sant'Anna, Pedro H.C.},
  year     = {2021},
  pages    = {S0304407620303948}
}

@article{Wooldridge_2025,
  title={Two-way fixed effects, the two-way mundlak regression, and difference-in-differences estimators},
  author={Wooldridge, Jeffrey M},
  journal={Empirical Economics},
  pages={1--43},
  year={2025},
  publisher={Springer}
}

@article{Roth_2022,
  title={Pretest with caution: Event-study estimates after testing for parallel trends},
  author={Roth, Jonathan},
  journal={American Economic Review: Insights},
  volume={4},
  number={3},
  pages={305--322},
  year={2022},
  publisher={American Economic Association 2014 Broadway, Suite 305, Nashville, TN 37203}
}

@article{Wang_et_al_2019,
  title={On the asymptotic distribution of model averaging based on information criterion},
  author={Wang, Miaomiao and Zhang, Xinyu and Wan, Alan TK and Zou, Guohua},
  journal={arXiv preprint arXiv:1910.12208},
  year={2019}
}

@article{Zhang_Liu_2019_inference,
  title={Inference after model averaging in linear regression models},
  author={Zhang, Xinyu and Liu, Chu-An},
  journal={Econometric Theory},
  volume={35},
  number={4},
  pages={816--841},
  year={2019},
  publisher={Cambridge University Press}
}

@article{Leeb_Potscher_2005,
  title={Model selection and inference: Facts and fiction},
  author={Leeb, Hannes and P{\"o}tscher, Benedikt M},
  journal={Econometric Theory},
  volume={21},
  number={1},
  pages={21--59},
  year={2005},
  publisher={Cambridge University Press}
}

@article{Kitagawa_Muris_2016,
  title={Model averaging in semiparametric estimation of treatment effects},
  author={Kitagawa, Toru and Muris, Chris},
  journal={Journal of Econometrics},
  volume={193},
  number={1},
  pages={271--289},
  year={2016},
  publisher={Elsevier}
}

@article{Steel_2020,
  title={Model averaging and its use in economics},
  author={Steel, Mark FJ},
  journal={Journal of Economic Literature},
  volume={58},
  number={3},
  pages={644--719},
  year={2020},
  publisher={American Economic Association 2014 Broadway, Suite 305, Nashville, TN 37203-2425}
}

@article{Sun_Abraham_2021,
  title={Estimating dynamic treatment effects in event studies with heterogeneous treatment effects},
  author={Sun, Liyang and Abraham, Sarah},
  journal={Journal of Econometrics},
  volume={225},
  number={2},
  pages={175--199},
  year={2021},
  publisher={Elsevier}
}

@article{goodman-bacon_2021,
  title={Difference-in-differences with variation in treatment timing},
  author={Goodman-Bacon, Andrew},
  journal={Journal of Econometrics},
  volume={225},
  number={2},
  pages={254--277},
  year={2021},
  publisher={Elsevier}
}

@article{Borusyak_et_al_2024,
  title={Revisiting event-study designs: robust and efficient estimation},
  author={Borusyak, Kirill and Jaravel, Xavier and Spiess, Jann},
  journal={Review of Economic Studies},
  volume={91},
  number={6},
  pages={3253--3285},
  year={2024},
  publisher={Oxford University Press UK}
}

@article{Hansen_2007,
  title={Least squares model averaging},
  author={Hansen, Bruce E},
  journal={Econometrica},
  volume={75},
  number={4},
  pages={1175--1189},
  year={2007},
  publisher={Wiley Online Library}
}

@article{Hansen_Racine_2012,
  title={Jackknife model averaging},
  author={Hansen, Bruce E and Racine, Jeffrey S},
  journal={Journal of Econometrics},
  volume={167},
  number={1},
  pages={38--46},
  year={2012},
  publisher={Elsevier}
}

@techreport{Firpo_et_al_2025,
  title={Model Averaging in Semiparametric Estimation of Quantile Treatment Effects},
  author={Firpo, Sergio and Galvao, Antonio F and Hounyo, Ulrich and Lu, Li},
  year={2025}
}

@article{Gobillon_Magnac_2016,
  title    = {Regional Policy Evaluation: Interactive Fixed Effects and Synthetic Controls},
  volume   = {98},
  journal  = {Review of Economics and Statistics},
  author   = {Gobillon, Laurent and Magnac, Thierry},
  year     = {2016},
  pages    = {535--551}
}

@article{Xu_2017,
  title    = {Generalized Synthetic Control Method: Causal Inference with Interactive Fixed Effects Models},
  volume   = {25},
  journal  = {Political Analysis},
  author   = {Xu, Yiqing},
  year     = {2017},
  pages    = {57--76}
}

@article{Callaway_Karami_2023,
  title={Treatment effects in interactive fixed effects models with a small number of time periods},
  author={Callaway, Brantly and Karami, Sonia},
  journal={Journal of Econometrics},
  volume={233},
  number={1},
  pages={184--208},
  year={2023},
  publisher={Elsevier}
}

@article{Arkhangelsky_et_al_2021,
	title        = {Synthetic difference-in-differences},
	author       = {Arkhangelsky, Dmitry and Athey, Susan and Hirshberg, David A and Imbens, Guido W and Wager, Stefan},
	year         = 2021,
	journal      = {American Economic Review},
	volume       = 111,
	pages        = {4088--4118}
}

@book{Anderson_1984, 
 title={An Introduction to Multivariate Statistical Analysis},
 author={Anderson, Theodore W.},
 year={1984},
 publisher={Wiley}
}

@article{Lu_2015,
  title={A covariate selection criterion for estimation of treatment effects},
  author={Lu, Xun},
  journal={Journal of Business \& Economic Statistics},
  volume={33},
  number={4},
  pages={506--522},
  year={2015},
  publisher={Taylor \& Francis}
}

@article{Gale_et_al_2024,
  title={Sweeping changes and an uncertain legacy: The Tax Cuts and Jobs Act of 2017},
  author={Gale, William G and Hoopes, Jeffrey L and Pomerleau, Kyle},
  journal={Journal of Economic Perspectives},
  volume={38},
  number={3},
  pages={3--32},
  year={2024},
  publisher={American Economic Association 2014 Broadway, Suite 305, Nashville, TN 37203-2418}
}

@incollection{Neumark_Simpson_2015,
  title={Place-based policies},
  author={Neumark, David and Simpson, Helen},
  booktitle={Handbook of regional and urban economics},
  volume={5},
  pages={1197--1287},
  year={2015},
  publisher={Elsevier}
}

@article{Busso_Gregory_Kline_2013,
  title={Assessing the incidence and efficiency of a prominent place based policy},
  author={Busso, Matias and Gregory, Jesse and Kline, Patrick},
  journal={American Economic Review},
  volume={103},
  number={2},
  pages={897--947},
  year={2013},
  publisher={American Economic Association}
}

@article{Ham_et_al_2011,
  title={Government programs can improve local labor markets: Evidence from state enterprise zones, federal empowerment zones and federal enterprise community},
  author={Ham, John C and Swenson, Charles and {\.I}mrohoro{\u{g}}lu, Ay{\c{s}}e and Song, Heonjae},
  journal={Journal of Public Economics},
  volume={95},
  number={7-8},
  pages={779--797},
  year={2011},
  publisher={Elsevier}
}

@article{Neumark_Young_2019,
  title={Enterprise zones, poverty, and labor market outcomes: Resolving conflicting evidence},
  author={Neumark, David and Young, Timothy},
  journal={Regional Science and Urban Economics},
  volume={78},
  pages={103462},
  year={2019},
  publisher={Elsevier}
}

@article{Freedman_Khanna_Neumark_2023,
  title={JUE insight: The impacts of opportunity zones on zone residents},
  author={Freedman, Matthew and Khanna, Shantanu and Neumark, David},
  journal={Journal of Urban Economics},
  volume={133},
  pages={103407},
  year={2023},
  publisher={Elsevier}
}

@article{Chan_Kwok_2022,
  title={The PCDID approach: difference-in-differences when trends are potentially unparallel and stochastic},
  author={Chan, Marc K and Kwok, Simon S},
  journal={Journal of Business \& Economic Statistics},
  volume={40},
  number={3},
  pages={1216--1233},
  year={2022},
  publisher={Taylor \& Francis}
}

@article{Callaway_Tsyawo_2023,
  title={Treatment effects in staggered adoption designs with non-parallel trends},
  author={Callaway, Brantly and Tsyawo, Emmanuel Selorm},
  journal={arXiv preprint arXiv:2308.02899},
  year={2023}
}

@article{Athey_et_al_2026,
  title={Triply Robust Panel Estimators},
  author={Athey, Susan and Imbens, Guido and Qu, Zhaonan and Viviano, Davide},
  journal={Journal of Applied Econometrics},
  year={2026},
  publisher={Wiley Online Library}
}

@article{Schenk_2025,
  title={Time-Weighted Difference-in-Differences in Short Panels with Common Shocks},
  author={Schenk, Timo},
  year={2025}
}

@article{Dench_et_al_2024,
  title={The effects of post-Dobbs abortion bans on fertility},
  author={Dench, Daniel and Pineda-Torres, Mayra and Myers, Caitlin},
  journal={Journal of Public Economics},
  volume={234},
  pages={105124},
  year={2024},
  publisher={Elsevier}
}

@online{YouGov,
  title = {Most Americans don’t want Roe v. Wade overturned, but it was widely expected even before the leak},
  author = {{YouGov PLC}},
  year = {2022},
  url = {https://yougov.com/en-us/articles/42385-likelihood-roe-v-wade-overturned-scotus-leak},
  urldate = {2026-07-14}
}

@article{Politis_Romano_1994,
  title={Large sample confidence regions based on subsamples under minimal assumptions},
  author={Politis, Dimitris N and Romano, Joseph P},
  journal={The Annals of Statistics},
  pages={2031--2050},
  year={1994},
  publisher={JSTOR}
}

@article{Politis_Romano_2010,
  title={K-sample subsampling in general spaces: The case of independent time series},
  author={Politis, Dimitris N and Romano, Joseph P},
  journal={Journal of multivariate analysis},
  volume={101},
  number={2},
  pages={316--326},
  year={2010},
  publisher={Elsevier}
}

@article{Buckland_et_al_1997,
  title={Model selection: an integral part of inference},
  author={Buckland, Steven T and Burnham, Kenneth P and Augustin, Nicole H},
  journal={Biometrics},
  pages={603--618},
  year={1997},
  publisher={JSTOR}
}

@article{Hjort_Claeskens_2003,
  title={Frequentist model average estimators},
  author={Hjort, Nils Lid and Claeskens, Gerda},
  journal={Journal of the American Statistical Association},
  volume={98},
  number={464},
  pages={879--899},
  year={2003},
  publisher={Taylor \& Francis}
}

@article{Wan_Zhang_Zou_2010,
  title={Least squares model averaging by Mallows criterion},
  author={Wan, Alan TK and Zhang, Xinyu and Zou, Guohua},
  journal={Journal of Econometrics},
  volume={156},
  number={2},
  pages={277--283},
  year={2010},
  publisher={Elsevier}
}

@article{Leeb_Potscher_2008,
  title={Can one estimate the unconditional distribution of post-model-selection estimators?},
  author={Leeb, Hannes and P{\"o}tscher, Benedikt M},
  journal={Econometric Theory},
  volume={24},
  number={2},
  pages={338--376},
  year={2008},
  publisher={Cambridge University Press}
}

\newpage~\section*{Appendix}

\subsection*{Proof of Lemma 3.1}

Let $\bm x_i\equiv(1,D_i)'$, $\bm\lambda_{s,t}\equiv(\gamma_{s,t},\tau_{s,t})'$, and
\begin{equation*}
    \bm Q_N\equiv\frac{1}{N}\isum\bm x_i\bm x_i',
    \qquad
    \bm g_{N,s,t}\equiv\frac{1}{N}\isum\bm x_i e_{i,s,t}.
\end{equation*}
On the event that both groups are non-empty, the OLS normal equations give
$\widehat{\bm\lambda}_{s,t}-\bm\lambda_{s,t}=\bm Q_N^{-1}\bm g_{N,s,t}$ and the exact identity
\begin{equation*}
    \frac{1}{N}\widehat{\SSR}_{s,t}
    =
    \frac{1}{N}\SSR_{s,t}^*
    -
    \bm g_{N,s,t}'\bm Q_N^{-1}\bm g_{N,s,t}.
\end{equation*}
By the definition of the population projection, $\expec{\bm x_i e_{i,s,t}}=\bm0$. The CLT gives $\bm g_{N,s,t}=O_p(N^{-1/2})$, and the WLLN gives $\bm Q_N\plim\expec{\bm x_i\bm x_i'}$, which is positive definite because $p\in(0,1)$. Hence
\begin{equation}\label{eq: SSR asymptotic equivalence}
    \frac{1}{N}\widehat{\SSR}_{s,t}
    =
    \frac{1}{N}\SSR_{s,t}^*+O_p(N^{-1}).
\end{equation}
Finite fourth moments imply $Var(e_{i,s,t}^2)<\infty$, so another application of the CLT yields
$N^{-1}\SSR_{s,t}^*=\rho_{s,t}+O_p(N^{-1/2})$.
Since $e_{i,s,t}$ is the long difference centered at its group-specific mean,
\begin{equation*}
    \rho_{s,t}
    =
    p Var(y_{i,s}-y_{i,t}\mid D_i=1)
    +(1-p)Var(y_{i,s}-y_{i,t}\mid D_i=0),
\end{equation*}
which proves the result.
\hfill $\square$

\subsection*{Proof of Lemma 3.2}

Fix $r\in\T_s^1$ and $t\in\T_s^2$. Lemma \ref{lemma:SSR}, equation \eqref{eq:SSRT-S}, and the separation condition imply
$\Delta_{s,t,r}\equiv\log(\rho_{s,t}/\rho_{s,r})>0$.
With probability approaching one,
$\log(\widehat{\SSR}_{s,t}/\widehat{\SSR}_{s,r})\geq\Delta_{s,t,r}/2$.
Since $\widehat w_{s,r}\leq1$, on this event
\begin{equation*}
    \widehat w_{s,t}
    \leq
    \frac{\widehat w_{s,t}}{\widehat w_{s,r}}
    =
    \exp\left\{-\frac{\sqrt N}{2}
    \log\left(\frac{\widehat{\SSR}_{s,t}}{\widehat{\SSR}_{s,r}}\right)\right\}
    \leq
    \exp(-\sqrt N\Delta_{s,t,r}/4).
\end{equation*}
The bound is $o(N^{-1/2})$, so $\widehat w_{s,t}=o_p(N^{-1/2})$.
\hfill $\square$

\subsection*{Proof of Lemma 3.3}

The positive-gap argument in Lemma \ref{lemma: unconditional bad model weights} also gives
$\widehat w_{s,t}=o_p(N^{-1/2})$ for $t\in\T_s^1\setminus\T_s^{1,*}$.
For $r\in\T_s^{1,*}$, equation \eqref{eq: SSR asymptotic equivalence} gives
\begin{equation*}
    Z_{N,s,r}\equiv
    \frac{\sqrt N(\widehat{\SSR}_{s,r}/N-\rho_s^*)}{\rho_s^*}
    =
    \frac{1}{\sqrt N}\isum\xi_{i,s,r}+o_p(1).
\end{equation*}
Stacking over the retained candidates and applying the multivariate CLT yields
$\bm Z_{N,s}\convd\bm Z_s\sim N(\bm0,\bm\Xi_s)$.
A Taylor expansion of the logarithm gives, jointly over these candidates,
\begin{equation*}
    \sqrt N\{\log(\widehat{\SSR}_{s,r}/N)-\log(\rho_s^*)\}
    =
    Z_{N,s,r}+o_p(1).
\end{equation*}
Subtracting the common term $\sqrt N\log(\rho_s^*)$ from the criteria does not change the weights. For every candidate outside $\T_s^{1,*}$, the remaining logarithmic gap has a strictly positive limit, so its exponential contribution vanishes. The retained exponential terms and the reciprocal of their sum are $O_p(1)$ because $\bm Z_{N,s}$ is tight. Consequently,
\begin{equation*}
    \widehat w_{s,r}
    =
    \frac{\exp(-Z_{N,s,r}/2)}
    {\sum_{q\in\T_s^{1,*}}\exp(-Z_{N,s,q}/2)}
    +o_p(1),\qquad r\in\T_s^{1,*}.
\end{equation*}
The continuous mapping theorem proves the stated joint weight limit.
\hfill $\square$

\subsection*{Proof of Theorem 3.1}

Let $\widehat p\equiv N_1/N$, $\bm\mu_{d,s}\equiv\condexpec{\bm W_s\bm y_i}{D_i=d}$,
$\bm\tau_s\equiv\bm\mu_{1,s}-\bm\mu_{0,s}$, and
$\widehat{\bm\tau}_s\equiv(\widehat\tau_{s,1},\ldots,\widehat\tau_{s,T_0})'$.
The difference of group means satisfies
\begin{equation*}
    \widehat{\bm\tau}_s-\bm\tau_s
    =
    \frac{1}{N}\isum
    \left\{
    \frac{D_i}{\widehat p}(\bm W_s\bm y_i-\bm\mu_{1,s})
    -
    \frac{1-D_i}{1-\widehat p}(\bm W_s\bm y_i-\bm\mu_{0,s})
    \right\}.
\end{equation*}
Since $\widehat p-p=O_p(N^{-1/2})$ and the group-centered sums are $O_p(\sqrt N)$,
\begin{equation*}
    \sqrt N(\widehat{\bm\tau}_s-\bm\tau_s)
    =
    \frac{1}{\sqrt N}\isum\bm\psi_{i,s}+O_p(N^{-1/2}).
\end{equation*}
Together with Lemma \ref{lemma: unconditional good model weights}, the fixed-dimensional CLT gives
\begin{equation*}
    \begin{pmatrix}
        (\sqrt N(\widehat{\bm\tau}_s-\bm\tau_s))_{s\in\mathcal S}\\
        (\bm Z_{N,s})_{s\in\mathcal S}
    \end{pmatrix}
    =
    \frac{1}{\sqrt N}\isum\bm V_i+o_p(1)
    \convd
    \begin{pmatrix}\bm\eta\\\bm Z\end{pmatrix}
    \sim N(\bm0,\bm\Sigma).
\end{equation*}
The weight expansion in that lemma and continuous mapping therefore yield
\begin{equation*}
    \left(
    \sqrt N(\widehat{\bm\tau}_s-\bm\tau_s),\widehat{\bm w}_s
    \right)_{s\in\mathcal S}
    \convd
    (\bm\eta_s,\bm w_s^*)_{s\in\mathcal S}.
\end{equation*}
For each $s$, $\sum_t\widehat w_{s,t}=1$ and $\tau_{s,r}=\tau_s$ for $r\in\T_s^{1,*}$. Lemmas \ref{lemma: unconditional bad model weights} and \ref{lemma: unconditional good model weights} give
\begin{equation*}
    \sqrt N\sum_{t\notin\T_s^{1,*}}
    \widehat w_{s,t}(\tau_{s,t}-\tau_s)=o_p(1).
\end{equation*}
Consequently,
\begin{align*}
    \left(
    \sqrt N\{\widehat\tau_s(\widehat{\bm w}_s)-\tau_s\}
    \right)_{s\in\mathcal S}
    &=
    \left(
    \widehat{\bm w}_s'\sqrt N(\widehat{\bm\tau}_s-\bm\tau_s)
    \right)_{s\in\mathcal S}+o_p(1)\\
    &\convd
    (\bm w_s^{*\prime}\bm\eta_s)_{s\in\mathcal S}.
\end{align*}
\hfill $\square$

\subsection*{Proof of Lemma 4.1}

Only the rows of $\bm C_s$ indexed by $\T_s^{1,*}$ contribute to $\bm w_s^{*\prime}\bm C_s$. For $t,r\in\T_s^{1,*}$, the influence-function definitions and $\condexpec{e_{i,s,t}}{D_i}=0$ give
\begin{equation*}
    [\bm C_s]_{t,r}
    =
    \frac{\condexpec{e_{i,s,t}e_{i,s,r}^2}{D_i=1}
    -\condexpec{e_{i,s,t}e_{i,s,r}^2}{D_i=0}}{\rho_s^*}.
\end{equation*}
Let $A_i\equiv\tau_{i,s}-\tau_s$ and $B_{i,t}\equiv u_{i,s}-u_{i,t}$.
Factor matching implies $e_{i,s,t}=D_iA_i+B_{i,t}$.
For treated units,
\begin{align*}
    &\condexpec{(A_i+B_{i,t})(A_i+B_{i,r})^2}{D_i=1}\\
    ={}&\condexpec{A_i^3}{D_i=1}
    +2\condexpec{A_i^2B_{i,r}}{D_i=1}
    +\condexpec{A_iB_{i,r}^2}{D_i=1}\\
    &+\condexpec{A_i^2B_{i,t}}{D_i=1}
    +2\condexpec{A_iB_{i,t}B_{i,r}}{D_i=1}
    +\condexpec{B_{i,t}B_{i,r}^2}{D_i=1}.
\end{align*}
Condition (ii) and the zero conditional means of $A_i$ and $B_{i,q}$ eliminate the four mixed terms. Condition (iii) eliminates the first term, and condition (i) makes the last term equal to
$\condexpec{B_{i,t}B_{i,r}^2}{D_i=0}$.
Thus $[\bm C_s]_{t,r}=0$ on the retained rows and $\bm w_s^{*\prime}\bm C_s=\bm0$; other rows need not vanish.

The retained subvector $\bm\eta_s^{**}$ is therefore independent of $\bm Z_s$ by joint Gaussianity. Conditional on $\bm Z_s$,
\begin{equation*}
    \frac{\bm w_s^{*\prime}\bm\eta_s}
    {\sqrt{\bm w_s^{*\prime}\bm\Omega_s\bm w_s^*}}
    \sim N(0,1),
\end{equation*}
so the unconditional law is also standard normal.
Consistency of the group means and treatment share, together with finite second moments of the long differences, gives
$N^{-1}\isum\norm{\widehat{\bm\psi}_{i,s}-\bm\psi_{i,s}}^2=o_p(1)$.
The WLLN and Cauchy--Schwarz then yield $\widehat{\bm\Omega}_s\plim\bm\Omega_s$.
Moreover,
\begin{equation*}
    \bm w_s^{*\prime}\bm\Omega_s\bm w_s^*
    \geq
    \frac{\lambda_{\min}(\bm\Omega_s^{**})}{|\T_s^{1,*}|}>0,
\end{equation*}
because the retained weights lie in the unit simplex.
The joint limit in Theorem \ref{theorem: unconditional asymptotic distribution} and continuous mapping therefore give the stated studentized limit.
\hfill $\square$

\subsection*{Proof of Lemma \ref{lemma:selector}}

The finite candidate set, Lemma \ref{lemma:SSR}, and the separation condition imply $\widehat\rho_s\plim\rho_s^*>0$. With probability approaching one, its minimum is attained in $\T_s^{1,*}$. For these candidates, the joint CLT and the delta method give
\begin{equation*}
    \max_{r\in\T_s^{1,*}}
    |\log(\widehat{\SSR}_{s,r}/N)-\log(\rho_s^*)|
    =O_p(N^{-1/2}).
\end{equation*}
Taking the minimum preserves this order, so
\begin{equation*}
    \log(\widehat\rho_s)-\log(\rho_s^*)=O_p(N^{-1/2}).
\end{equation*}
Consequently, for $r\in\T_s^{1,*}$,
\begin{equation*}
    a_N\{\log(\widehat{\SSR}_{s,r}/N)-\log(\widehat\rho_s)\}
    =O_p(a_N/\sqrt N)=o_p(1),
\end{equation*}
and $\widehat s_{s,r}(a_N)\plim1$. For $t\notin\T_s^{1,*}$, the logarithmic gap converges to $\log(\rho_{s,t}/\rho_s^*)>0$. Its product with $a_N$ diverges to $+\infty$ in probability, hence $\widehat s_{s,t}(a_N)\plim0$. The convergence is joint over the finite set of candidates and post-treatment periods.

\hfill $\square$

\subsection*{Proof of Theorem \ref{theorem:subsampling}}

Fix $s$ and write $\widehat\tau_s\equiv\widehat\tau_s(\widehat{\bm w}_s)$. Theorem \ref{theorem: unconditional asymptotic distribution} gives
\begin{equation*}
    \sqrt N(\widehat\tau_s-\tau_s)\convd
    L_s\equiv\bm w_s^{*\prime}\bm\eta_s.
\end{equation*}
We show that the subsampling distribution estimates the distribution of this limiting root, then establish quantile consistency and coverage.

\noindent\emph{Uniform unit subsampling.}
For a continuity point $x$ of $F_s$, define
\begin{gather*}
    h_{N,x}(I)\equiv
    \mathds{1}\{\sqrt{b_N}(\widehat\tau_{s,I}-\tau_s)\leq x\},
    \qquad
    L_N^0(x)\equiv
    \binom{N}{b_N}^{-1}\sum_{|I|=b_N}h_{N,x}(I),
\end{gather*}
where $\widehat\tau_{s,I}$ is MADID recomputed on $I$, including the weights with $\sqrt{b_N}$ scaling. Thus $L_N^0$ uses every possible subset and centers at the true ATT. Each fixed subset contains $b_N$ i.i.d. units, so
\begin{equation*}
    \expec{L_N^0(x)}
    =\Pr\{\sqrt{b_N}(\widehat\tau_{s,b_N}-\tau_s)\leq x\}
    \longrightarrow F_s(x)
\end{equation*}
by Theorem \ref{theorem: unconditional asymptotic distribution} and $b_N\rightarrow\infty$.

The overlapping subset indicators are not independent. To bound the variance of their average, let $\mathcal D_N$ denote the observed panel and randomly permute the units. Divide the first $k_Nb_N$ units into $k_N=\lfloor N/b_N\rfloor$ disjoint blocks, leaving the remaining units unused, and set
\begin{equation*}
    A_\pi(x)\equiv
    \frac{1}{k_N}\sum_{\ell=1}^{k_N}h_{N,x}(I_{\pi,\ell}).
\end{equation*}
Each block is a uniform $b_N$-unit subset conditional on the panel, giving $\condexpec{A_\pi(x)}{\mathcal D_N}=L_N^0(x)$. For any fixed permutation, its disjoint block indicators are independent under i.i.d. sampling, each with variance at most $1/4$ and the same expectation. The law of total variance therefore yields
\begin{align*}
    Var(L_N^0(x))
    &=Var\{\condexpec{A_\pi(x)}{\mathcal D_N}\}\\
    &\leq Var(A_\pi(x))
    \leq\frac{1}{4k_N}=O(b_N/N)\longrightarrow0.
\end{align*}
Consequently,
\begin{equation*}
    \expec{(L_N^0(x)-F_s(x))^2}
    =Var(L_N^0(x))
    +\{\expec{L_N^0(x)}-F_s(x)\}^2
    \longrightarrow0,
\end{equation*}
and $L_N^0(x)\plim F_s(x)$.

\noindent\emph{Estimated centering and random subsets.}
The implemented root replaces the unknown $\tau_s$ by $\widehat\tau_s$. This changes the ideal root by
\begin{equation*}
    c_N\equiv\sqrt{b_N}(\widehat\tau_s-\tau_s)
    =\sqrt{b_N/N}\,\sqrt N(\widehat\tau_s-\tau_s)
    =O_p(\sqrt{b_N/N})=o_p(1).
\end{equation*}
Thus the distribution over all subsets with estimated centering is $L_N(x)\equiv L_N^0(x+c_N)$. On $|c_N|\leq\varepsilon$,
\begin{equation*}
    L_N^0(x-\varepsilon)\leq L_N(x)\leq L_N^0(x+\varepsilon).
\end{equation*}
Apply the preceding convergence at $x\pm\varepsilon$ and let $\varepsilon\downarrow0$ through continuity points. Continuity at $x$ gives $L_N(x)\plim F_s(x)$. This step uses $b_N/N\rightarrow0$, not a zero-mean restriction on $L_s$.

Conditional on the panel, the $B$ random subsets give independent Bernoulli indicators with mean $L_N(x)$. Hence
\begin{gather*}
    \condexpec{\widehat L_{N,B}(x)}{\mathcal D_N}=L_N(x),
    \qquad
    \condexpec{(\widehat L_{N,B}(x)-L_N(x))^2}{\mathcal D_N}
    \leq\frac{1}{4B}\longrightarrow0.
\end{gather*}
It follows that $\widehat L_{N,B}(x)\plim F_s(x)$.

\noindent\emph{Fixed treated and control allocations.}
For the $K$-sample construction of \citet{Politis_Romano_2010}, we must also check that fixing the group counts preserves the MADID root limit. The possible difference is that i.i.d. sampling includes treatment-share fluctuations, whereas fixed allocations suppress them.

Under \eqref{eq:subsampling group moments}, $\rho_s^*=pm_{s,1}+(1-p)m_{s,0}$. For $r\in\T_s^{1,*}$, decompose
\begin{gather*}
    \xi_{i,s,r}
    =
    \xi_{i,s,r}^{\circ}
    +\frac{(D_i-p)(m_{s,1}-m_{s,0})}{\rho_s^*},
    \qquad
    \xi_{i,s,r}^{\circ}\equiv
    \frac{e_{i,s,r}^2-m_{s,D_i}}{\rho_s^*},\\
    \condexpec{\xi_{i,s,r}^{\circ}}{D_i}=0,
    \qquad
    \condexpec{\bm\psi_{i,s}}{D_i}=\bm0.
\end{gather*}
The first term is a within-group centered fluctuation. The second is the treatment-share contribution; it is identical across retained candidates. Its Gaussian limit therefore adds a common offset to their criteria, which cancels from the weights:
\begin{equation*}
    \frac{\exp\{-(z_r+c)/2\}}
    {\sum_{q\in\T_s^{1,*}}\exp\{-(z_q+c)/2\}}
    =
    \frac{\exp(-z_r/2)}
    {\sum_{q\in\T_s^{1,*}}\exp(-z_q/2)}.
\end{equation*}
Thus the root depends only on the joint fluctuations of $\bm\psi_{i,s}$ and the retained $\xi_{i,s,r}^{\circ}$, not on the omitted common offset. Because these influence functions have conditional mean zero, their i.i.d. covariance is the mixture of their within-group covariances with weights $p$ and $1-p$. The fixed-allocation two-sample CLT gives this same covariance when $b_{1,N}/b_N\plim p$. The allocation-rate condition also ensures that first-order allocation drift vanishes:
\begin{align*}
    \sqrt{b_N}(b_{1,N}/b_N-p)
    &=
    \sqrt{b_N}(b_{1,N}/b_N-N_1/N)
    +\sqrt{b_N/N}\sqrt N(N_1/N-p)\\
    &=o_p(1).
\end{align*}
The joint influence-function expansion and weight map from Theorem \ref{theorem: unconditional asymptotic distribution} therefore give the same root limit $L_s$ for independent samples of sizes $b_{1,N}$ and $b_{0,N}$ from the two conditional populations.

Let $L_N^{0,K}(x)$ be the distribution over all such treated--control subset pairs, centered at $\tau_s$. Conditional on the labels $D_1,\ldots,D_N$, observations within each group are i.i.d. and the groups are independent. Its conditional expectation is the root distribution in the preceding two-sample experiment. Randomly partition each group into disjoint blocks and pair $k_N^K\equiv\min_d\lfloor N_d/b_{d,N}\rfloor$ of them. The same conditional-averaging and variance argument as above gives
\begin{align*}
    \condexpec{L_N^{0,K}(x)}{D_1,\ldots,D_N}
    &\plim F_s(x),\\
    Var(L_N^{0,K}(x)\mid D_1,\ldots,D_N)
    &\leq\frac{1}{4k_N^K}\\
    &=O_p\left(\frac{b_{0,N}}{N_0}+\frac{b_{1,N}}{N_1}\right)=o_p(1).
\end{align*}
Conditional Chebyshev's inequality yields $L_N^{0,K}(x)\plim F_s(x)$. The estimated-centering and Monte Carlo arguments then apply without change.

\noindent\emph{Quantiles and coverage.}
Let $q_\beta\equiv F_s^{-1}(\beta)$ for $\beta\in\{\alpha/2,1-\alpha/2\}$. Continuity and strict increase at $q_\beta$ imply, for sufficiently small $\varepsilon>0$ with continuity points $q_\beta\pm\varepsilon$,
\begin{equation*}
    F_s(q_\beta-\varepsilon)<\beta<F_s(q_\beta+\varepsilon).
\end{equation*}
Convergence of the empirical distribution at these two points gives
\begin{equation*}
    \Pr\{\widehat L_{N,B}(q_\beta-\varepsilon)<\beta
    <\widehat L_{N,B}(q_\beta+\varepsilon)\}\longrightarrow1.
\end{equation*}
On this event, $q_\beta-\varepsilon<\widehat q_\beta\leq q_\beta+\varepsilon$, so $\widehat q_\beta\plim q_\beta$. Combining this with the full-sample root limit,
\begin{align*}
    &\Pr\left\{
    \widehat\tau_s-\frac{\widehat q_{1-\alpha/2}}{\sqrt N}
    \leq\tau_s\leq
    \widehat\tau_s-\frac{\widehat q_{\alpha/2}}{\sqrt N}
    \right\}\\
    &\qquad=
    \Pr\{\widehat q_{\alpha/2}
    \leq\sqrt N(\widehat\tau_s-\tau_s)
    \leq\widehat q_{1-\alpha/2}\}\\
    &\qquad\longrightarrow
    \Pr\{q_{\alpha/2}\leq L_s\leq q_{1-\alpha/2}\}
    =1-\alpha.
\end{align*}
The distributional argument also applies to joint roots for any fixed collection of post-treatment periods. The result is pointwise and does not assert an optimal choice of $b_N$.
\hfill $\square$

\subsection*{Proof of Theorem \ref{theorem:simulated inference}}

Define
\begin{equation*}
    \widetilde\xi_{i,s,t}\equiv
    \frac{e_{i,s,t}^2-\rho_{s,t}}{\rho_{s,t}},
    \qquad t=1,\ldots,T_0,
\end{equation*}
and let $\widetilde{\bm V}_i$ stack all DID and residual-square influence functions in the order of $\widehat{\bm V}_i$.
Put $\delta_{N,s,t}=|\widehat\gamma_{s,t}-\gamma_{s,t}|+|\widehat\tau_{s,t}-\tau_{s,t}|=O_p(N^{-1/2})$. Then
\begin{align*}
    |\widehat e_{i,s,t}-e_{i,s,t}|&\leq\delta_{N,s,t},\\
    \frac{1}{N}\isum(\widehat e_{i,s,t}^2-e_{i,s,t}^2)^2
    &\leq
    8\delta_{N,s,t}^2\frac{1}{N}\isum e_{i,s,t}^2
    +2\delta_{N,s,t}^4=o_p(1).
\end{align*}
Finite fourth moments, consistency of the positive variance denominators, the WLLN, and Cauchy--Schwarz give
\begin{gather*}
    \frac{1}{N}\isum\norm{\widehat{\bm V}_i-\widetilde{\bm V}_i}^2=o_p(1),
    \qquad
    \widehat{\bm\Sigma}\plim
    \widetilde{\bm\Sigma}\equiv
    \expec{\widetilde{\bm V}_i\widetilde{\bm V}_i'}.
\end{gather*}
The covariance block retaining all DID components and the criterion components in $\T_s^{1,*}$ equals $\bm\Sigma$.

Let $\widehat{\bm s}\equiv(\widehat s_{s,t}(a_N))_{s,t}$ and $\bm s^*\equiv(\mathds{1}(t\in\T_s^{1,*}))_{s,t}$. Lemma \ref{lemma:selector} and continuity of the positive semidefinite square root imply
\begin{equation*}
    \widehat{\bm s}\plim\bm s^*,
    \qquad
    \widehat{\bm\Sigma}^{1/2}\plim\widetilde{\bm\Sigma}^{1/2}.
\end{equation*}
For $\bm v=(\bm\eta',\bm Z')'$, define the simulation map
\begin{equation*}
    H_s(\bm v,\bm a)\equiv
    \frac{\sum_{t=1}^{T_0}a_{s,t}\exp(-Z_{s,t}/2)\eta_{s,t}}
    {\sum_{t=1}^{T_0}a_{s,t}\exp(-Z_{s,t}/2)}.
\end{equation*}
Its denominator at $\bm a=\bm s^*$ is positive for every finite $\bm Z$. For an independent standard Gaussian vector $\bm g$ of the appropriate dimension,
\begin{equation*}
    \left(
    H_s(\widetilde{\bm\Sigma}^{1/2}\bm g,\bm s^*)
    \right)_{s\in\mathcal S}
    \ \stackrel{d}{=}\
    (L_s)_{s\in\mathcal S}.
\end{equation*}
Conditional continuous mapping therefore gives, for every bounded continuous $f$,
\begin{equation*}
    \condexpec{f((\widehat\tau_s^{(1)})_{s\in\mathcal S})}{\mathcal D_N}
    \plim
    \expec{f((L_s)_{s\in\mathcal S})},
\end{equation*}
where $\mathcal D_N$ is the observed panel. Conditional independence of the draws yields
\begin{equation*}
    Var\left(
    \frac{1}{B}\sum_{b=1}^B
    f((\widehat\tau_s^{(b)})_{s\in\mathcal S})
    \,\middle|\,\mathcal D_N
    \right)
    \leq\frac{\norm{f}_\infty^2}{B}\longrightarrow0.
\end{equation*}
Quantile consistency and coverage follow as in the proof of Theorem \ref{theorem:subsampling}, without recentering the simulated roots.
\hfill $\square$

\clearpage
\section*{Additional Simulation Results}
\renewcommand{\thetable}{A.\arabic{table}}
\setcounter{table}{0}

Table~\ref{tab:simulation N} varies the sample size at $\sigma_u = 1$. Bias and RMSE decline with $N$ in the sinusoidal, manual, and constant-factor designs. Under complete misspecification, the linear-trend weights increasingly concentrate on the best-fitting invalid candidate; the resulting bias persists while coverage deteriorates.

\begin{table}[htbp]
\centering
\caption{MADID across sample sizes ($\sigma_u=1$)}
\label{tab:simulation N}
\small
\begin{tabular}{lrrrrr}
\toprule
Design & $N$ & Bias & SD & RMSE & CP \\
\midrule
Sinusoidal & 50 & -0.099 & 0.348 & 0.362 & 0.973 \\
 & 200 & -0.012 & 0.168 & 0.168 & 0.964 \\
 & 1000 & 0.005 & 0.084 & 0.084 & 0.912 \\
\midrule
Linear trend & 50 & 0.362 & 0.399 & 0.538 & 0.921 \\
 & 200 & 0.251 & 0.200 & 0.321 & 0.861 \\
 & 1000 & 0.216 & 0.089 & 0.234 & 0.468 \\
\midrule
Manual & 50 & -0.080 & 0.341 & 0.350 & 0.968 \\
 & 200 & -0.024 & 0.169 & 0.171 & 0.948 \\
 & 1000 & 0.001 & 0.074 & 0.074 & 0.945 \\
\midrule
Constant factor & 50 & 0.007 & 0.319 & 0.319 & 0.970 \\
 & 200 & -0.005 & 0.159 & 0.159 & 0.956 \\
 & 1000 & 0.002 & 0.069 & 0.069 & 0.954 \\
\bottomrule
\end{tabular}
\end{table}

Table~\ref{tab:simulation covariance} considers Gaussian idiosyncratic errors with $Corr(u_{i,t},u_{i,s})=c_u^{|t-s|}$ under the manual factor path, $N=200$, and $\sigma_u=1$. The differenced-error variance is $2\sigma_u^2\{1-c_u^{|t-s|}\}$, so serial correlation changes the candidate SSRs differently according to their distance from the post-treatment period.

\begin{table}[htbp]
\centering
\caption{MADID under AR(1) idiosyncratic errors: manual factor path}
\label{tab:simulation covariance}
\small
\begin{tabular}{rrrrr}
\toprule
$c_u$ & Bias & SD & RMSE & CP \\
\midrule
0.0 & -0.024 & 0.169 & 0.171 & 0.948 \\
0.3 & -0.048 & 0.167 & 0.174 & 0.947 \\
0.6 & -0.044 & 0.155 & 0.161 & 0.933 \\
0.9 & -0.000 & 0.086 & 0.086 & 0.958 \\
\bottomrule
\end{tabular}
\end{table}

Table~\ref{tab:simulation weights} reports average MADID weights. In the sinusoidal and manual designs, the weights increasingly concentrate on valid periods. In the linear-trend design, they concentrate on period 5, the closest available comparison to the post-treatment factor. In the constant-factor design, the weights remain approximately uniform because all candidate criteria are equivalent.

\begin{table}[htbp]
\centering
\caption{MADID candidate weights ($\sigma_u=1$)}
\label{tab:simulation weights}
\small
\begin{tabular}{lrrrrrr}
\toprule
Design & $N$ & $\overline w_1$ & $\overline w_2$ & $\overline w_3$ & $\overline w_4$ & $\overline w_5$ \\
\midrule
Sinusoidal & 50 & 0.051 & 0.425 & 0.050 & 0.425 & 0.049 \\
 & 200 & 0.006 & 0.492 & 0.006 & 0.489 & 0.006 \\
 & 1000 & 0.000 & 0.495 & 0.000 & 0.505 & 0.000 \\
\midrule
Linear trend & 50 & 0.018 & 0.046 & 0.119 & 0.283 & 0.535 \\
 & 200 & 0.001 & 0.005 & 0.035 & 0.209 & 0.751 \\
 & 1000 & 0.000 & 0.000 & 0.001 & 0.056 & 0.944 \\
\midrule
Manual & 50 & 0.279 & 0.283 & 0.075 & 0.289 & 0.074 \\
 & 200 & 0.319 & 0.320 & 0.021 & 0.319 & 0.021 \\
 & 1000 & 0.331 & 0.343 & 0.001 & 0.324 & 0.001 \\
\midrule
Constant factor & 50 & 0.200 & 0.199 & 0.204 & 0.200 & 0.198 \\
 & 200 & 0.196 & 0.202 & 0.199 & 0.196 & 0.206 \\
 & 1000 & 0.205 & 0.198 & 0.199 & 0.197 & 0.201 \\
\bottomrule
\end{tabular}
\end{table}

\end{document}